\documentclass[12pt]{article}

\usepackage[utf8]{inputenc}
\usepackage{textcomp}
\usepackage{amsmath}
\usepackage[shortlabels]{enumitem}
\usepackage{chetnew}
\usepackage{tikz}
\usepackage{amssymb,amsfonts,mathrsfs,dsfont,yfonts,bbm}\usetikzlibrary{calc}
\usepackage{xcolor}
\usepackage{braket}
\usepackage[normalem]{ulem}
\usepackage{booktabs}
\usepackage{pst-node}
\usepackage{tikz-cd} 

\newcommand{\be}{\begin{equation}}
\newcommand{\ee}{\end{equation}}
\newcommand{\bea}{\begin{eqnarray}}
\newcommand{\eea}{\end{eqnarray}}

\newcommand{\CA}{\mathcal{A}}

\newcommand{\CD}{\mathcal{D}}
\newcommand{\CE}{\mathcal{E}}

\newcommand{\CB}{\mathcal{B}}
\newcommand{\CC}{\mathcal{C}}

\newcommand{\CO}{\mathcal{O}}
\newcommand{\CT}{\mathcal{T}}

\newcommand{\CN}{\mathcal{N}}
\newcommand{\CS}{\mathcal{S}}

\newcommand*{\boxcoloro}{orange}
\makeatletter
\newcommand{\boxedo}[1]{\textcolor{\boxcoloro}{%
\tikz[baseline={([yshift=-1ex]current bounding box.center)}] \node [rectangle, minimum width=1ex,rounded corners,draw] {\normalcolor\m@th$\displaystyle#1$};}}
 \makeatother
\newcommand*{\boxcolorr}{red}
\makeatletter
\newcommand{\boxedr}[1]{\textcolor{\boxcolorr}{%
\tikz[baseline={([yshift=-1ex]current bounding box.center)}] \node [rectangle, minimum width=1ex,rounded corners,draw] {\normalcolor\m@th$\displaystyle#1$};}}
 \makeatother
\newcommand*{\boxcolorb}{blue}
\makeatletter
\newcommand{\boxedb}[1]{\textcolor{\boxcolorb}{%
\tikz[baseline={([yshift=-1ex]current bounding box.center)}] \node [rectangle, minimum width=1ex,rounded corners,draw] {\normalcolor\m@th$\displaystyle#1$};}}
 \makeatother
\newcommand*{\boxcolorg}{green}
\makeatletter
\newcommand{\boxedg}[1]{\textcolor{\boxcolorg}{%
\tikz[baseline={([yshift=-1ex]current bounding box.center)}] \node [rectangle, minimum width=1ex,rounded corners,draw] {\normalcolor\m@th$\displaystyle#1$};}}
 \makeatother
 \newcommand*{\boxcolorp}{purple}
\makeatletter
\newcommand{\boxedp}[1]{\textcolor{\boxcolorp}{%
\tikz[baseline={([yshift=-1ex]current bounding box.center)}] \node [rectangle, minimum width=1ex,rounded corners,draw] {\normalcolor\m@th$\displaystyle#1$};}}
 \makeatother
  \newcommand*{\boxcolorc}{cyan}
\makeatletter
\newcommand{\boxedc}[1]{\textcolor{\boxcolorc}{%
\tikz[baseline={([yshift=-1ex]current bounding box.center)}] \node [rectangle, minimum width=1ex,rounded corners,draw] {\normalcolor\m@th$\displaystyle#1$};}}
 \makeatother
  \newcommand*{\boxcolory}{yellow}
\makeatletter
\newcommand{\boxedy}[1]{\textcolor{\boxcolory}{%
\tikz[baseline={([yshift=-1ex]current bounding box.center)}] \node [rectangle, minimum width=1ex,rounded corners,draw] {\normalcolor\m@th$\displaystyle#1$};}}
 \makeatother

\usepackage{amsmath}	
\begin{document}

\title{Non-Perturbative Explorations of \\[3mm] Chiral Rings in 4d $\CN=2$ SCFTs}

\author{Anindya Banerjee$^{\spadesuit}$ and Matthew Buican$^{\diamondsuit}$}

\affiliation{\smallskip CTP and Department of Physics and Astronomy\\
Queen Mary University of London, London E1 4NS, UK\emails{$^{\spadesuit}$a.banerjee@qmul.ac.uk, $^{\diamondsuit}$m.buican@qmul.ac.uk}}

\abstract{We study the conditions under which 4d $\CN=2$ superconformal field theories (SCFTs) have multiplets housing operators that are chiral with respect to an $\CN=1$ subalgebra. Our main focus is on the set of often-ignored and relatively poorly understood $\bar\CB$ representations. These multiplets typically evade direct detection by the most popular non-perturbative 4d $\CN=2$ tools and correspondences. In spite of this fact, we demonstrate the ubiquity of $\bar\CB$ multiplets and show they are associated with interesting phenomena. For example, we give a purely algebraic proof that they are present in all local unitary $\CN>2$ SCFTs. We also show that $\bar\CB$ multiplets exist in $\CN=2$ theories with rank greater than one and a conformal manifold or a freely generated Coulomb branch. Using recent topological quantum field theory results, we argue that certain $\bar\CB$ multiplets exist in broad classes of theories with the $\mathbb{Z}_2$-valued 't Hooft-Witten anomaly for $Sp(N)$ global symmetry. Motivated by these statements, we then study the question of when $\bar\CB$ multiplets exist in rank-one SCFTs with exactly $\CN=2$ SUSY and vanishing 't Hooft-Witten anomaly. We conclude with various open questions.}

\medskip

\date{June 2023}
\setcounter{tocdepth}{2}
\maketitle

\tableofcontents 
\newsec{Introduction}
Chiral operators play a fundamental role in 4D $\CN=2$ quantum field theories (QFTs) at all length scales. At short distances, the allowed $\CN=2$-preserving relevant deformations of $\CN=2$ superconformal field theories (SCFTs) are chiral \cite{Argyres:2015ffa,Cordova:2016xhm}. Expectation values of chiral operators parameterize the $\CN=2$ moduli spaces of these SCFTs and initiate renormalization group (RG) flows to vacua where low-energy vector multiplets and hypermultiplets live. These effective multiplets are also chiral.\footnote{In this paper, unless otherwise specified, an operator is termed \lq\lq chiral" if it is chiral with respect to some $\CN=1\subset\CN=2$ subalgebra and is non-trivial in the corresponding ring (see section \ref{Bbargen} for more details). A multiplet is considered chiral if it houses a chiral operator.}

Since chiral operators are so ubiquitous, it is no surprise that many of the most important non-perturbative insights into 4d $\CN=2$ QFTs are intimately connected with these operators. For example, Seiberg-Witten geometries \cite{Seiberg:1994rs,Seiberg:1994aj} encode the exact infrared (IR) prepotential, which is a chiral object. Higgs branches enjoy various non-renormalization theorems \cite{Argyres:1996eh}, and their associated chiral operators are closely related to 2d VOAs \cite{Beem:2013sza} and hidden infinite-dimensional symmetries.

Given their prominence and the powerful geometrical and algebraic constraints on their spectra, one may have the impression that chiral sectors of $\CN=2$ theories are completely understood, simple to characterize, and probe phenomena that are well known. Each of these statements is far from the truth.

To better understand these points, it is helpful to first think about ultraviolet (UV) physics and understand which superconformal representations house chiral operators. As we will review in the next section, this question was answered in \cite{Bhargava:2022cuf}. A basic but important point is that the operators directly connected with the relatively well-understood Coulomb branch physics of Seiberg-Witten theory and the physics of the Higgs branch sit in half-BPS multiplets.\footnote{Even these sectors are not fully understood in general. For example, it is believed (without proof) that any interacting 4d $\CN=2$ SCFT has a Coulomb branch. But even basic properties of the Coulomb branch, such as the most general conditions under which its corresponding chiral ring is freely generated, are not known.} In the nomenclature of \cite{Dolan:2002zh}, these are the $\bar\CE$ and $\hat\CB$ multiplets (a more detailed discussion appears in the next section).\footnote{More precisely, $\hat\CB$ multiplets are chiral with respect to half the supersymmetry and anti-chiral with respect to the other half. The remaining multiplets housing chiral operators (except for $\bar\CE$) satisfy less restrictive shortening conditions. See the next section for further details.} In particular, vevs for the superconformal primaries (SCPs) of these multiplets parameterize the Coulomb and Higgs branches respectively.

However, $\CN=2$ SCFTs contain less protected multiplets with chiral primaries, and these multiplets give rise to interesting physics. For example, $\bar\CD$ multiplets are also intimately connected with 2d VOAs \cite{Beem:2013sza} and probe various subtle properties of the topology and punctures of class $\CS$ compactification surfaces (e.g., see \cite{Kang:2022zsl}). Moreover, some $\bar\CD$ multiplets contain the extra supercurrents of $\CN>2$ SCFTs while others capture the physics of the Weinberg-Witten theorem. 

Still, from a purely algebraic point of view, the above multiplets are not the most general representations containing chiral operators. Indeed, while these multiplets have primaries that depend on at most two quantum numbers, there are more general multiplets with primaries that depend on three quantum numbers. These are the $\bar\CB$ multiplets and are the main focus of this paper.

Given the greater freedom in their quantum numbers, one might wonder if $\bar\CB$ multiplets are ubiquitous in 4d $\CN=2$ SCFTs. Unfortunately, the answer to this question is obscured by the fact that these multiplets are under less control then the $\bar\CE$, $\hat\CB$, and $\bar\CD$ representations. For example, $\bar\CB$ multiplets are not captured by any of the special limits of the superconformal index. Moreover, Seiberg-Witten theory and the 4d/2d correspondence of \cite{Beem:2013sza} do not directly detect these degrees of freedom.\footnote{Seiberg-Witten curves indirectly detect certain $\bar\CB$ multiplets in the low energy description of the Coulomb branch.}

One well-known instance where $\bar\CB$ multiplets appear is whenever a UV theory has a \lq\lq mixed" branch. This is a branch of moduli space where low-energy vector multiplets and hypermultiplets co-exist. In other words, mixed branches include a Coulomb branch and a Higgs branch component at common points in moduli space. Therefore, we expect a $\bar\CB$ multiplet to appear in the following operator product
\begin{equation}
\bar\CE\times\hat\CB\ni \bar\CB~,
\end{equation}
where the $\bar\CB$ primary is a composite built out of a Coulomb branch $\bar\CE$ primary and a Higgs branch $\hat\CB$ primary. Giving an expectation value to the $\bar\CB$ primary gives a vev to both $\bar\CE$ and $\hat\CB$ primaries and initiates an RG flow to a mixed branch. However, in many theories, null relations set $\bar\CB=0$ and lead to geometrically separate Coulomb and Higgs branches.

The main purpose of this paper is to explain a much broader array of phenomena that are captured by $\bar\CB$ multiplets beyond the existence of a mixed branch. Indeed, we will argue that
\begin{itemize}
\item All local unitary 4d $\CN>2$ SCFTs have $\bar\CB$ multiplets. We give an algebraic proof of this fact that follows purely from locality and unitarity (see Section \ref{N3susy}).
\item All higher-rank 4d $\CN=2$ SCFTs with conformal manifolds parameterized by gauge couplings\footnote{All known examples of 4d $\CN=2$ conformal manifolds have a gauge coupling interpretation. Such families of SCFTs generally have \lq\lq matter" sectors that are interacting isolated SCFTs (as opposed to only containing collections of free hypermultiplets whose symmetries are gauged).} have $\bar\CB$ multiplets that exist at all points on the conformal manifold\footnote{It is straightforward to construct $\bar\CB$ multiplets that exist at special points on the conformal manifold (or, more generally, for special values of a gauge coupling). For example, at zero gauge coupling we can construct, for $SU(N)$ (and $N>2$), $\bar\CB$ primaries of the schematic form ${\rm Tr}\phi^2\CO$, where $\CO$ is a $\hat\CB_R$ primary transforming in the adjoint of $SU(N)$, and $\phi$ is the corresponding vector multiplet scalar. However, such operators are not protected from recombination and typically become part of long multiplets as we turn on the gauge coupling (here we have taken the generic case $R>1/2$; in the special case of a free matter sector with $R=1/2$, we do obtain a protected multiplet). Our interest is in $\bar\CB$ multiplets that are robust against quantum corrections, are present everywhere on the conformal manifold, and do not require considering special matter sectors.\label{gauging}} (see Section \ref{higher-rank}).
\item All higher-rank 4d $\CN=2$ SCFTs with freely generated Coulomb branches have $\bar\CB$ multiplets (see Section \ref{higher-rank}).
\item Any 4d $\CN=2$ SCFT with an $Sp(N)$ symmetry having a $\mathbb{Z}_2$-valued 't Hooft anomaly \cite{Witten:1982fp} has a $\bar\CB$ multiplet if its Coulomb branch has at least one point consisting of purely free fields (see Section \ref{Wanomsec}).
\end{itemize}
Therefore, we will see that $\bar\CB$ multiplets are indeed ubiquitous and that they are related to various interesting phenomena.

Given the above results, we are also motivated to study rank-one SCFTs with purely $\CN=2$ SUSY, no $\mathbb{Z}_2$-valued 't Hooft anomaly, and no mixed branch. Indeed, the existence of $\bar\CB$ multiplets in these theories is not implied by the above results. For example, in the case of the rank-one $\CN=2$ theory studied in \cite{Bhargava:2022cuf}, such multiplets were shown to be absent. The main tool used in that paper was $\CN=2$ superconformal representation theory coupled with the dynamics of $\CN=1\to\CN=2$ SUSY enhancement along an RG flow to the IR. Using similar techniques, we will study various other rank-one theories amenable to such analysis. In all such isolated theories that we study (for simplicity, we stick to those of Argyres-Douglas type), we will see that $\bar\CB$ multiplets are absent.

The plan of the paper is as follows. In the next section, we introduce various details of the superconformal analysis of chiral operators in 4d $\CN=2$ SCFTs. In section \ref{genresults} we present the general results described in the bullet points above. We are then motivated to make some conjectures on the spectrum of $\bar\CB$ multiplets in general theories. In the remainder of the paper, we study various rank-one SCFTs and conclude with a discussion of future directions. 

\newsec{$\bar{\CB}$ multiplets and superconformal representation theory}\label{Bbargen}
In this section, we briefly discuss the superconformal representation theory and ring structure of $\CN=2$ multiplets that contain an operator that is chiral with respect to an $\CN=1\subset\CN=2$ subalgebra. We conclude by explaining where $\bar\CB$ multiplets sit in this universe.

Recall that an $\CN=2$ superconformal field theory has an $SU(2)_R\times U(1)_R$ symmetry with eight Poincar\'e and eight special supercharges transforming as doublets under the $R$ symmetry. Without loss of generality, we follow \cite{Bhargava:2022cuf} and take our $\CN=1$ subalgebra to be generated by the following Poincar\'e supercharges
\begin{equation}\label{N1alg}
Q_{1\alpha}\sim Q^2_{\alpha}\in\left(  \frac12,0\right)_{-\frac12,-\frac12}~,\ \ \ \bar Q^1_{\dot\alpha}\sim\bar Q_{2\dot\alpha}\in\left(0,  \frac12\right)_{\frac12,\frac12}~,
\end{equation}
where the quantum numbers are $(j,\bar j)_{R,r}$, with $(j,\bar j)$ the left and right spin, $R$ the $SU(2)_R$ weight, and $r$ the $U(1)_r$ charge (note that, to get a superconformal subalgebra, we should also include the special supercharges $S_{2\alpha}\sim S^1_{\alpha}$ and $\bar S_{1\dot\alpha}\sim\bar S^2_{\dot\alpha}$).

With these conventions, the $\CN=1$ chiral operators are those satisfying
\begin{equation}\label{chirdef}
\left[\bar Q^1_{\dot\alpha},\CO\right\}=0~,\ \ \ \CO\ne\left\{\bar Q^1_{\dot\alpha},\CO'\right]~.
\end{equation}
Such operators form a chiral ring (their OPEs are free from singularities), and the second condition in \eqref{chirdef} is equivalent to demanding that $\CO$ is non-trivial in this ring (here $\CO'$ is any well-defined local operator in the theory). Note that we have suppressed any $SU(2)_R$ and Lorentz quantum numbers of $\CO$.

In the context of $\CN=2$ SCFTs, operators satisfying \eqref{chirdef} can sit in various representations. We find homes for our chiral operators in these multiplets by acting on chiral superconformal primaries with the (chiral part) of the subalgebra orthogonal to \eqref{N1alg} \cite{Bhargava:2022cuf}
\be\label{N1ortho}
Q_{2\alpha}\sim Q^1_{\alpha}\in\left(  \frac12,0\right)_{ \frac12,-\frac12}~,\ \ \ \bar Q^2_{\dot\alpha}\sim\bar Q_{1\dot\alpha}\in\left( 0,  \frac12\right)_{-\frac12, \frac12}~.
\ee  
The analysis of \cite{Bhargava:2022cuf} shows that operators satisfying \eqref{chirdef} can only sit in the following positions in an $\CN=2$ multiplet
\bea \label{Chiralloc}
\CO^{11\cdots1}_{\alpha_1\cdots\alpha_{2j}}\in (j,0)_{R,r} &\xrightarrow{Q^1_\alpha}& \CO_{\alpha_1\cdots\alpha_{2j}\alpha}^{11\cdots11}\oplus\CO_{\alpha_1\cdots\alpha_{2j-1}\alpha}^{11\cdots11}\in \left(j+\frac12,0\right)_{R+\frac12,r-\frac12}\oplus\left(j-\frac12,0\right)_{R+\frac12,r-\frac12} \cr&\xrightarrow{(Q^1)^2 }&  \CO_{\alpha_1\cdots\alpha_{2j}}^{11\cdots111}\in(j,0)_{R+1,r-1}~,
\eea 
Here the leftmost operator is a chiral superconformal primary of an $\CN=2$ multiplet (it has highest $SU(2)_R$ weight), and the remaining operators are successive $Q^1_{\alpha}$ descendants. Depending on the multiplet in question, some of the descendants may be null.

In the language of \cite{Dolan:2002zh}, solutions to \eqref{Chiralloc} are exhausted by multiplets in the so-called \lq\lq full chiral sector" (FCS) \cite{Bhargava:2022cuf}
\begin{equation}\label{FCSdef}
{\rm FCS}:=\bar\CE_{r}\oplus\hat\CB_R\oplus \bar\CD_{R(j,0)}\oplus\bar\CB_{R,r(j,0)}~.
\end{equation}
Let us analyse these multiplets in turn:
\begin{itemize}
\item{The $\bar\CE_{r}$ primary is $U(1)_r$ charged but has $R=j=\bar j=0$ \cite{Manenti:2019jds}. It is annihilated by all the $\bar Q^i_{\alpha}$. In this sense, it is maximally protected. According to the standard lore, it is also the most universal FCS multiplet: such multiplets exist in all known interacting 4d $\CN=2$ SCFTs, and their vevs give coordinates on the Coulomb branch of these SCFTs. The superconformal primaries form the closed Coulomb branch chiral (sub)ring.\footnote{Such operators therefore give rise to coordinates in Seiberg-Witten geometries and their generalizations. This fact explains their ubiquity, although it is not completely clear to us if their ubiquity is also a consequence of the type of 4d $\CN=2$ SCFTs we have been able to construct to date. Ideally, one would like to understand if such multiplets emerge from some more minimal set of algebraic criteria.} More generally, these multiplets house three $\CN=1$ chiral operators from \eqref{Chiralloc}
\bea \label{Eloc}
\CO\in (0,0)_{0,r} &\xrightarrow{Q^1_\alpha}& \CO_{\alpha}^{1}\in \left({1\over2},0\right)_{1/2,r-1/2}\xrightarrow{(Q^1)^2 }  \CO^{11}\in(0,0)_{1,r-1}~.
\eea
The descendant states do not form part of the Coulomb branch chiral ring.
}
\item{The $\hat\CB_R$ multiplets have $r=j=\bar j=0$ and $R>0$. All known examples of these multiplets parameterize Higgs branches of $\CN=2$ SCFTs via the vevs of their primaries. They form a closed Higgs branch chiral (sub)ring. Therefore, while these multiplets are also common, they are more special than the $\bar\CE$ type. Indeed, the SCFT in question has to have sufficient \lq\lq matter" for a Higgs branch to exist. These multiplets form Virasoro primaries under the 4d/2d map of \cite{Beem:2013sza} and hence are under good analytic control. $\hat\CB_R$ multiplets house only one $\CN=1$ chiral operator (the primary)
\bea \label{Bloc}
\CO\in (0,0)_{R,0} &\xrightarrow{Q^1_\alpha}& 0~.
\eea 
As a result, the primary is also anti-chiral with respect to the orthogonal algebra in \eqref{N1ortho}. Like the $\bar\CE_r$ multiplet, $\hat\CB_R$ is therefore maximally protected under the SUSY algebra (although the primary is not annihilated by all of the same supercharges).
\item{The $\bar\CD_{R(j,0)}$ multiplet has $\bar j=0$ and $r=1+j$. In general, it has $R,j>0$, but it can also have $j=0$ and $R\ge0$. This multiplet is ubiquitous in low energy effective theories: the $\bar\CD_{0(0,0)}$ multiplet contains the chiral half of the free vector and is therefore present on the Coulomb branch of any theory. More generally, $\bar\CD$ multiplets with $R>0$ appear if we also have decoupled hypermultiplets (or more complicated matter sectors with Higgs branches) appearing on the Coulomb branch.\footnote{Such a phenomenon indicates the presence of a mixed branch.} Indeed, consider the following IR OPE 
\begin{equation}\label{DBOPE}
\bar\CD_{0,(0,0)}\times\hat\CB_R\ni\bar\CD_{R(0,0)}~.
\end{equation}
By studying the OPE of primaries (since there are no singularities, this is just the chiral ring product), we see that a $\bar\CD_{R(0,0)}$ multiplet must appear on the righthand side.

The $\bar\CD$ multiplet is less ubiquitous in interacting theories. This relative scarcity is because the $\bar\CD$ $U(1)_r$ charge is fixed in terms of the multiplet's spin to be at a unitarity bound. Moreover, recall that in flows to the Coulomb branch, $U(1)_r$ is broken but Lorentz symmetry is not. Therefore, it is common for $\bar\CD$ multiplets to arise from UV multiplets with larger $U(1)_r$ charge like $\bar\CE$ or $\bar\CB$ multiplets.

However, $\bar\CD$ multiplets always appear in local (interacting) theories with $\CN>2$ SUSY since the $\bar\CD_{1/2(0,0)}$ multiplet houses the extra SUSY currents. More generally, $\bar\CD$ multiplets appear in certain interacting $\CN=2$ theories as well. For example, they exist in class $\CS$ theories whose corresponding Riemann surfaces have non-trivial $\pi_1$ (for more general examples, see \cite{Kang:2022zsl}). These multiplets therefore seem to know interesting things about topology. They also give rise to Virasoro primaries under the 4d/2d map of \cite{Beem:2013sza} and are therefore under stringent analytic control. For generic $R$ and $j$, they house the following $\CN=1$ chiral primaries
\bea \label{ChirallocD}
\CO^{11\cdots1}_{\alpha_1\cdots\alpha_{2j}}\in (j,0)_{R,j+1} &\xrightarrow{Q^1_\alpha}& \CO_{\alpha_1\cdots\alpha_{2j}\alpha}^{11\cdots11}\in \left(j+\frac12,0\right)_{R+\frac12,j+\frac12} \xrightarrow{(Q^1)^2 }0~.
\eea 
}
\item{Finally, we consider the $\bar\CB_{R,r(j,0)}$ multiplets of interest. They are clearly the most general FCS multiplets in the sense that they have three independent quantum numbers $R>0$, $j$, and $r>1+j$ (only $\bar j=0$). Moreover, all states in \eqref{Chiralloc} are present
\bea \label{Chiralloc2}
\CO^{11\cdots1}_{\alpha_1\cdots\alpha_{2j}}\in (j,0)_{R,r} &\xrightarrow{Q^1_\alpha}& \CO_{\alpha_1\cdots\alpha_{2j}\alpha}^{11\cdots11}\oplus\CO_{\alpha_1\cdots\alpha_{2j-1}\alpha}^{11\cdots11}\in \left(j+\frac12,0\right)_{R+\frac12,r-\frac12}\oplus\left(j-\frac12,0\right)_{R+\frac12,r-\frac12} \cr&\xrightarrow{(Q^1)^2 }&  \CO_{\alpha_1\cdots\alpha_{2j}}^{11\cdots111}\in(j,0)_{R+1,r-1}~.
\eea 

On a mixed branch, these multiplets are as common as $\bar\CD$ multiplets. For example, in the presence of free vectors, we have $\bar\CE_n$ operators from the $n$-fold $\bar\CD_{0(0,0)}^{\times n}\ni\bar\CE_n$ OPE. We can then repeat the IR OPE in \eqref{DBOPE} but with $\bar\CD_{0(0,0)}\to\bar\CE_n$ and $\bar\CD_{R(0,0)}\to\bar\CB_{R,n(0,0)}$. We will show below that $\bar\CB$ multiplets exist whenever a theory has a freely generated Coulomb branch of rank at least two (in this sense they are slightly less ubiquitous than the $\bar\CD$ multiplets since they do not appear in the theory of a single free vector \cite{Bhargava:2022cuf}).

In interacting theories, we expect such multiplets to be more common than $\bar\CD$ multiplets. This is because $r>1+j$ is an inequality (as opposed to the equality in the $\bar\CD$ case). For example, we expect
\begin{equation}\label{BEOPE}
\bar\CE_r\times\hat\CB_R\ni\bar\CB_{R,r(0,0)}~,
\end{equation}
whenever the SCFT supports a mixed branch. The vevs of the corresponding $\bar\CB_{R,r(0,0)}$ chiral primaries parameterize these mixed branches.

We can also find other chiral ring products giving rise to $\bar\CB_{R,r(j,0)}$ multiplets.\footnote{See \cite{Ramirez:2016lyk} for $\bar\CB$ production channels outside the chiral ring OPE (and footnote \ref{gauging} for production channels that do not involve OPEs of bulk local operators). We will not discuss these channels in this paper.} For example, selection rules allow
\begin{equation}\label{DDOPE}
\bar\CD_{R(j,0)}\times\bar\CD_{R'(j',0)}\ni\bar\CB_{R+R',j+j'+2(j+j',0)}~.
\end{equation}
Unless all the $\bar\CD$ primaries are minimally nilpotent in the chiral ring, they must give rise to corresponding $\bar\CB$ multiplets.\footnote{Indeed, in the next section, we will use algebraic techniques to show that the $\bar\CD_{1/2(0,0)}$ multiplets housing extra $\CN>2$ supercurrents are never minimally nilpotent.} Combined with \eqref{BEOPE}, this observation again suggests that $\bar\CB$ multiplets should be more common than $\bar\CD$ multiplets in interacting theories.

We can also imagine constructing $\bar\CB$ multiplets via chiral ring products involving descendants of the multiplets discussed in previous bullets.\footnote{In some cases this is impossible. For example,
\begin{eqnarray}
\bar\CD_{R(j,0)}\times\bar\CD_{R'(j',0)}&\ni& \CO_{\alpha_1\cdots\alpha_{2j}}^{1\cdots1}Q^1_{\alpha}\CO_{\alpha_1\cdots\alpha_{2j'}}^{'1\cdots1}+\kappa\CO_{\alpha_1\cdots\alpha_{2j'}}^{'1\cdots1}Q^1_{\alpha}\CO_{\alpha_1\cdots\alpha_{2j}}^{1\cdots1}\cr&\in&\bar\CB_{R+R'+1/2,j+j'+3/2(j+j'+1/2,0)}\cong\bar\CD_{R+R'+1/2(j+j'+1/2)}~,
\end{eqnarray}
where $\kappa\in\mathbb{C}$ is required to make the operator in question a superconformal primary. More generally, if we involve at most a single $\bar\CD_{R(j,0)}$ primary, we must also take spin contractions (this is because the descendant in \eqref{ChirallocD} has $r=j$).
}
For example, we can take
\begin{eqnarray}\label{EEdesc}
\bar\CE_{r}\times\bar\CE_{r'}&\ni& \CO Q^1_{\alpha}\CO'+\kappa Q^1_{\alpha}(\CO)\CO'\in\bar\CB_{1/2,r+r'-1/2(1/2,0)}~,\cr
\bar\CE_{r}\times\bar\CE_{r'}&\ni& Q^{1\alpha}\CO Q^1_{\alpha}\CO'+\kappa_1 (Q^1)^2(\CO)\CO'+\kappa_2\CO(Q^1)^2(\CO')\in\bar\CB_{1,r+r'-1(0,0)}~,\ \ \ \ \ \ 
\end{eqnarray}
where $\kappa,\kappa_1,\kappa_2\in\mathbb{C}$ are required to make the above operators superconformal primaries.\footnote{Note that if $\CO=\CO'$, then the $\bar\CB_{1/2,r+r'-1(1/2,0)}$ multiplet in \eqref{EEdesc} vanishes.}
}

More generally, it is apriori possible that $\bar\CB$ multiplets can appear as chiral ring generators.\footnote{Although, outside of theories involving very special matter sectors, the only such examples we are aware of are of the unprotected form discussed in footnote \ref{gauging}; one may also consider the possibility of obtaining such generators from gauging a non-anomalous discrete symmetry.}

Finally, we note that, at the level of multiplication of superconformal primaries in the chiral ring, $\bar\CB_{R,r(j,0)}$ multiplets form (two-sided) ideals
\begin{eqnarray}
\bar\CE_r\times\bar\CB_{R',r'(j',0)}&\ni&\bar\CB_{R',r+r'(j',0)}~,\  \hat\CB_R\times\bar\CB_{R',r'(j',0)}\ni\bar\CB_{R+R',r'(j',0)}~,\cr \bar\CD_{R(j,0)}\times\bar\CB_{R',r'(j',0)}&\ni&\bar\CB_{R+R',r'+j+1(j+j',0)}~,\cr \bar\CB_{R,r(j,0)}\times\bar\CB_{R',r'(j',0)}&\ni&\bar\CB_{R+R',r+r'(j+j',0)}~.
\end{eqnarray}
}
\end{itemize}

Therefore, to summarize: in the absence of FCS chiral ring relations, we expect $\bar\CB$ multiplets to be present whenever the theory is interacting (since we then expect $\bar\CE$ multiplets). Moreover, we expect the corresponding chiral primaries to form ideals in the chiral ring and therefore to be crucial in understanding the full local operator algebras of interacting 4d $\CN=2$ SCFTs.

However, $\CN=2$ theories often have FCS chiral ring relations\footnote{These relations need not involve only Coulomb branch primaries in general. Indeed, in the examples we discuss below, they do not.} (these relations will feature prominently in our rank-one discussion below), and so the above conclusion is too naive. Still, given how easy it is to generate such multiplets in the chiral ring, we expect $\bar\CB$ multiplets to be present in broad classes of theories and to detect various physical phenomena. Indeed, we will arrive at a few general results on these multiplets in the next section.

\newsec{General results}\label{genresults}
In this section, we discuss several abstract results on the presence of $\bar\CB$ multiplets in broad classes of 4d $\CN=2$ SCFTs. These results are connected with various physical phenomena.

\subsec{Local unitary SCFTs with $\CN\ge3$ SUSY}\label{N3susy}
Any local unitary SCFT with $\CN\ge3$ supersymmetry has a $\bar\CB$ multiplet.\footnote{This statement is highly non-trivial for local (non-Lagrangian) $\CN=3$ SCFTs. Note that it also applies to any potential (yet to be discovered) local non-Lagrangian $\CN=4$ theories.} Indeed, by locality, any such theory has an $\CN=3$ stress tensor multiplet. As a result, from an $\CN=2$ perspective, the theory has a $U(1)$ flavor symmetry (descending from the $\CN=3$ $R$ symmetry), which we will call $U(1)_G$. The Noether current for this symmetry sits in a corresponding $\hat \CB_1^{0}\cong B_1\bar B_1[0;0]^{(2;0),0}$ multiplet. Here we use the language of both \cite{Dolan:2002zh} and \cite{Cordova:2016emh}.\footnote{The additional superscript in $\hat \CB_1^{0}\cong B_1\bar B_1[0;0]^{(2;0),0}$ refers to the fact that this multiplet has zero $U(1)_G$ charge (we follow the conventions of \cite{Zafrir:2020epd}). We will only write this superscript explicitly in cases where the $U(1)_G$ charge is relevant to the argument.} The reason we introduce new nomenclature is that we will make a claim regarding the presence of $\bar\CB$ multiplets using $\CN=3$ superconformal representation theory, and \cite{Dolan:2002zh} only discusses $\CN=2$ representations.

The highest $SU(2)_R$-weight component of this $\hat \CB_1\cong B_1\bar B_1[0;0]^{(2;0)}$ multiplet is the holomorphic moment map, $M^{11}$. Together with the highest $SU(2)_R$-weight operator in the stress tensor multiplet, $\hat\CC_{0(0,0)}\cong A_2\bar A_{\bar2}[0;0]_2^{(0;0)}$, and in the extra supercurrent multiplets, $\CD_{{1\over2}(0,0)}\oplus\bar\CD_{{1\over2}(0,0)}\cong A_2\bar B_1[0,0]^{(1;2)}\oplus B_1\bar A_{2}[0;0]^{(1;-2)}$, $M^{11}$ is related by the chiral algebra map of \cite{Beem:2013sza} to the generators of a 2d $\CN=2$ super-Virasoro VOA \cite{Nishinaka:2016hbw}
\begin{equation}
\chi(M^{11})=J~,\ \ \ \chi(J^{11}_{+\dot+})=T~,\ \ \ \chi(J^{11}_+)=G~,\ \ \ \chi(J^{11}_{\dot+})=\bar G~.
\end{equation}
Here $J$ is a $U(1)$ affine current, $T$ is the 2d EM tensor, and the remaining operators are the 2d supercurrents (we refer the reader to \cite{Beem:2013sza,Nishinaka:2016hbw} for more detailed discussions of this correspondence).

We claim that in any local unitary $\CN\ge3$ theory, the 4d OPE of the holomorphic moment maps contains the following dimension-four $SU(2)_R$ weight-two operator
\begin{equation}\label{MMOPE}
M^{11}\times M^{11}\supset (M^{11})^2~,
\end{equation}
where $(M^{11})^2\in\hat\CB_2^0\cong B_1\bar B_1[0;0]^{(4;0),0}$. We can argue for this statement by recalling the following selection rules (e.g., see \cite{Kiyoshige:2018wol})\footnote{We only keep track of so-called \lq\lq Schur" multiplets in these selection rules. The reason is that these multiplets house the operators subject to the correspondence in \cite{Beem:2013sza}.}
\begin{equation}\label{selection}
\hat\CB_1\times\hat\CB_1=\hat\CB_1+\hat\CB_2+\sum_{\ell=0}^{\infty}\hat\CC_{0({\ell\over2},{\ell\over2})}+\sum_{\ell=0}^{\infty}\hat\CC_{1({\ell\over2},{\ell\over2})}~,
\end{equation}
and translating to the 2d VOA. Specializing to multiplets that can provide an $h=E-R=2$ operator in the OPE, it turns out that the righthand side of \eqref{selection} reduces to $\hat\CB_1+\hat\CB_2+\hat\CC_{0(0,0)}$. In the 2d picture, the $\hat\CB_1$ contribution to the OPE comes from $\chi(\partial M^{11})=\partial J$.

To verify \eqref{MMOPE}, we therefore need to check that there are no null relations involving $JJ$, $\partial J$, and $T$ (otherwise, the 4d normal-ordered product in \eqref{MMOPE} vanishes according to the general prescription in \cite{Beem:2013sza}). We can use the bosonic part of the super-Virasoro algebra
\begin{eqnarray}
[L_n,L_m]&=&(n-m)L_{n+m}+{c\over12}n(n^2-1)\delta_{n+m,0}~,\cr [J_n,J_m]&=&{c\over12}n\delta_{n+m,0}~,\cr [L_n,J_m]&=&-mJ_{n+m}~,
\end{eqnarray}
and compute the matrix of inner products
\begin{eqnarray}
\begin{pmatrix}
\langle T|T\rangle & \langle T|JJ\rangle & \langle T|\partial J\rangle\\ \langle JJ|T\rangle& \langle JJ|JJ\rangle& \langle JJ|\partial J\rangle\\ \langle \partial J| T\rangle& \langle \partial J|JJ\rangle& \langle \partial J|\partial J\rangle\end{pmatrix}&=&
\begin{pmatrix}
\langle L_2L_{-2}\rangle & \langle L_2J_{-1}J_{-1}\rangle & \langle L_2J_{-2}\rangle\\ \langle J_1J_1L_{-2}\rangle& \langle J_1J_1J_{-1}J_{-1}\rangle& \langle J_1J_1J_{-2}\rangle\\ \langle J_2 L_{-2}\rangle & \langle J_2J_{-1}J_{-1}\rangle & \langle J_2 J_{-2}\rangle\\ 
\end{pmatrix}\cr&=&\begin{pmatrix}
c/2 & c/12 & 0\\ c/12& c^2/72 & 0\\ 0 & 0 & c/6
\end{pmatrix}~.
\end{eqnarray}
The above matrix has determinant $c^3(c-1)/864$ and is therefore non-invertible for $c=0,1$. Under the 4d/2d map of \cite{Beem:2013sza}, these central charges map to 4d central charges $c_{4d}=0,-1/12$ and correspond to non-unitary 4d theories. Therefore, in a unitary 4d theory with $\CN\ge3$ SUSY, we see that we necessarily have a $\hat\CB_2\cong B_1\bar B_1[0;0]^{(4;0)}$ multiplet.

Where does the above multiplet sit in $\CN=3$ representation theory? A moment's thought indicates it must sit inside the $\CN=3$ stress-tensor multiplet self-OPE
\begin{equation}
B_1\bar B_1[0;0]^{(1,1;0)}\times B_1\bar B_1[0;0]^{(1,1;0)}\ni B_1\bar B_1[0;0]^{(4;0),0}\cong\hat\CB_2~.
\end{equation}
Note that the $B_1\bar B_1[0;0]^{(1,1;0)}$ multiplet transforms in the ${\bf8}$ of $SU(3)_R$. Now, recall that
\begin{equation}
{\bf8}\times{\bf8}={\bf1}+{\bf8}+{\bf8}+{\bf10}+{\bf{\overline{10}}}+{\bf27}~.
\end{equation}
Clearly, $B_1\bar B_1[0;0]^{(4;0),0}$ cannot sit in the first three $SU(3)_R$ representations above. Using branching rules, it is also easy to check that $(M^{11})^2$ transforms as part of a {\bf27} of $SU(3)_R$. In terms of $\CN=2$ $SU(2)_R$ representations, we have scaling dimension four Lorentz-scalar primaries
\begin{equation}\label{multdecomp}
{\bf27}={\bf1}+{\bf2}+{\bf2}+{\bf3}+{\bf3}+{\bf3}+{\bf4}+{\bf4}+{\bf5}~.
\end{equation}
These operators must have $U(1)_R^{\CN=3}$ charge zero and so the lefthand side is a representation of type $B_1\bar B_1[0;0]^{(2,2;0)}$. Therefore, on the righthand side the $U(1)_R^{\CN=2}$ charges of the primaries satisfy
\begin{equation}
U(1)_R^{\CN=2}={2\over3}U(1)_{SU(3)_R}=-2U(1)_G~.
\end{equation}
In terms of our conventions relative to \cite{Cordova:2016emh}, $U(1)_r:={1\over2}U(1)_R^{\CN=2}$.

Now we are ready to analyze the righthand side of \eqref{multdecomp}. The first representation must be a long multiplet (since $R=r=0$). The next two representations have $R=1/2$ and $r=1$ and are also long multiplets (they cannot be $\hat\CC$ multiplets since they have scalar primaries). The next three representations have $R=1$ and $r=2,0,-2$. This gives a $\bar\CB_{1,2(0,0)}\oplus\hat\CC_{1(0,0)}\oplus\CB_{1,-2(0,0)}$ triple.\footnote{Note that the $\hat \CC_{1(0,0)}$ multiplet is the universal multiplet described in \cite{Buican:2021elx} for an interacting theory with a flavor symmetry.} The next two representations have $R=3/2$ and $r=1,-1$. These are $\bar\CD_{3/2}(0,0)\oplus\CD_{-3/2(0,0)}$. The final representation is $\hat\CB_{2}$.

As a result, we have arrived at our promised statement: any local unitary $\CN\ge3$ SCFT has a $\bar\CB$ multiplet. In $\CN=2$ language, this multiplet can be understood as arising from the normal-ordered product of the extra supercurrent multiplet\footnote{Note that $\bar\CD_{1/2(0,0)}\times\CD_{1/2(0,0)}\ni \hat\CC_{1(0,0)}$.}
\begin{equation}
\bar\CD_{1/2(0,0)}\times\bar\CD_{1/2(0,0)}\ni \bar\CB_{1,2(0,0)}
\end{equation}
This is a particular example of the more general channel described in \eqref{DDOPE}.

Note that this discussion does not require the existence of a moduli space of vacua (although all known examples of $\CN\ge3$ theories have such moduli spaces). Moreover, although the 4d/2d VOA map of \cite{Beem:2013sza} doesn't directly detect $\bar\CB$ multiplets, we see that we can combine that map with locality and $\CN>2$ SUSY to deduce the existence of $\bar\CB$ multiplets.

In summary, we have the following result:

\bigskip
\noindent
{\bf Statement 1:} Any local unitary 4d $\CN\ge3$ SCFT has a $\bar\CB_{1,2(0,0)}$ multiplet.

\subsection{Higher-rank SCFTs}\label{higher-rank}
In the previous section, we saw that all local unitary $\CN\ge3$ SCFTs have $\bar\CB$ multiplets. However, these theories are special by virtue of their enhanced symmetry. It is natural to then wonder if $\bar\CB$ multiplets are always related to symmetry enhancement or other more special phenomena.

In this section, we will see the answer is no. In particular, we will demonstrate the ubiquity of $\bar\CB$ multiplets. Indeed, we will see that, under relatively relaxed assumptions, higher-rank theories have such multiplets.

Let us begin with a rank $N\ge2$ SCFT that is part of an $\CN=2$ or $\CN=4$ conformal manifold.\footnote{Recall that there are no $\CN=3$-preserving exactly marginal deformations.} All known exactly marginal deformations in $\CN=2$ SCFTs involve gauge couplings for some non-abelian group, $G$. Such theories always admit at least one weak gauge coupling limit in which the theory factorizes into a sector consisting of the vector multiplets, $V_G$, and one or more matter sectors, $\CT_i$.\footnote{The $\CT_i$ need not be weakly coupled themselves. For example, consider the Minahan-Nemeschansky ${ E_6}$ theory, ${\rm MN}_{E_6}$, appearing in the $SU(2)$ duality frame of \cite{Argyres:2007cn}.}  If the theory has rank $N\ge2$, this means that the we have at least two generators of the Coulomb branch chiral ring. One generator must be the quadratic Casimir, $\bar\CE_2\in V_G$, and the other generator is either a higher Casimir of $V_G$ (e.g., if $G=SU(3)$) or else is a Coulomb branch generator, $\bar\CE_r\in\CT_i$.\footnote{As an example of this latter phenomenon, consider the case of $\bar\CE_3\in {\rm MN}_{E_6}$ in the example of \cite{Argyres:2007cn}.} In general there is no invariant distinction between these two possibilities (e.g., as can be seen by looking at both sides of the duality in \cite{Argyres:2007cn}).

Let us now set the $G$ gauge coupling to zero. Then, we see that
\begin{equation}\label{Bhalfgen}
\CO_{2,r,\alpha}:=\CO_2Q^1_{\alpha}(\CO_r)+\kappa Q^1_{\alpha}(\CO_2) \CO_r~,
\end{equation}
is a superconformal primary for a particular choice of $\kappa\in\mathbb{C}$. Here $\CO_2$ is the $\bar\CE_2$ superconformal primary (it is related by supersymmetry to the exactly marginal deformation), and $\CO_r$ is some other Coulomb branch chiral ring generator. As a result, \eqref{Bhalfgen} is a primary of a $\bar\CB_{1/2,r+3/2(1/2,0)}$ multiplet. Since $\bar\CB_{1/2,r+3/2(1/2,0)}$ multiplets cannot recombine into long multiplets, this multiplet remains at {\it all} points on the conformal manifold (unlike generic operators of the form discussed in footnote \ref{gauging}). We therefore arrive at the following statement:

\bigskip
\noindent
{\bf Statement 2:} In any rank $N\ge 2$ 4d $\CN=2$ SCFT with an exactly marginal gauge coupling, there is at least one $r$ such that $\bar\CB_{1/2,r+3/2(1/2,0)}$ is in the spectrum.

\bigskip
Next let us consider rank $N\ge2$ SCFTs that are not necessarily part of a conformal manifold. For simplicity, let us assume that these theories have an $N\ge2$ dimensional Coulomb branch that is freely generated.

Then, in the free IR theory we flow to by turning on a vev for a chiral operator in the UV, we will encounter operators similar to those in \eqref{Bhalfgen}. For example, we have
\begin{equation}\label{BhalfgenIR}
\CO_{3,i,j\alpha}:=\phi_i\left(\phi_iQ^1_{\alpha}\phi_j-Q^1_{\alpha}(\phi_i) \phi_j\right)\in\bar\CB_{1/2,5/2(1/2,0)}~, \ \ \ i\ne j~,
\end{equation}
where $i,j=1,\cdots,N$ denote the particular free vector component in the $U(1)^N$ free super-Maxwell theory present at generic points on the Coulomb branch. For $N\ge2$ such $\bar\CB$ multiplets clearly exist, and this logic therefore explains the comment on higher-rank Coulomb branches below \eqref{Chiralloc2}.

What can the multiplets in \eqref{BhalfgenIR} descend from in the UV? Since superconformal recombination of $\bar\CB_{1/2,r(1/2,0)}$ multiplets is forbidden, it is tempting to argue that the multiplets in \eqref{BhalfgenIR} come from $\bar\CB_{1/2,r(1/2,0)}$ multiplets in the UV theory. However, we should be careful: in the flow back up to the UV we break superconformal (and $U(1)_r$) symmetry. Therefore, we should understand whether these multiplets can sit inside larger non-conformal multiplets.

To that end, let $\left\{\lambda_a\right\}$ denote the (generally infinite) collection of irrelevant couplings in the deformed IR theory that flows back up to the UV theory in question. Since supersymmetry and $SU(2)_R$ symmetry are both preserved, these couplings should sit as superconformal primaries in background / spurion multiplets with quantum numbers
\begin{equation}
R(\lambda_a)=j(\lambda_a)=\bar j(\lambda_a)=0~.
\end{equation}
If $\CO_{3,i,j,\alpha}$ becomes part of a longer non-conformal representation but remains chiral and a (SUSY) primary after introducing the $\lambda_a$, then, in the UV theory, we expect a corresponding $\bar\CB_{1/2,r(1/2,0)}$ multiplet when superconformal symmetry re-emerges. However, this scenario is incompatible with spontaneous superconformal symmetry breaking.

Instead, let us suppose that $\CO_{3,i,j,\alpha}$ is not a SUSY primary or is no longer chiral after turning on the $\lambda_a$. To that end, first suppose $\CO_{3,i,j,\alpha}$ is a SUSY primary satisfying
\begin{equation}\label{nc}
\bar Q^1_{\dot\alpha}\CO_{3,i,j,\alpha}=\lambda_a\CO_{3,i,j,\alpha\dot\alpha}\ne0~, \ \ \ (\bar Q^1)^2\CO_{3,i,j,\alpha}=0~.
\end{equation}
In the first equation in \eqref{nc} we could have considered a more general linear combination of operators multiplying different polynomials (series) in the $\lambda_a$. For simplicity, we have written a single such term with a single power of a coupling (but our arguments can be generalized straightforwardly to the most general case).

In the UV theory, the superconformal primary $\CO_{3,i,j,\alpha}$ satisfies the shortening condition in \eqref{nc}. It is therefore in either a $\bar\CB_{1/2,r(1/2,0)}$ or a $\bar\CC_{1/2,r(1/2,0)}$ multiplet (note that a $\bar\CD_{1/2,(1/2,0)}$ multiplet clearly contains too few degrees of freedom relative to the IR). We claim neither possibility is consistent.

To find the contradiction, let us first suppose that $\CO_{3,i,j,\alpha\dot\alpha}$ is an IR superconformal primary. Then it is a primary of an IR $\bar\CC_{1,r'(1/2,1/2)}$ or $\hat\CC_{1(1/2,1/2)}$ multiplet. However, such multiplets cannot sit inside a UV $\bar\CB_{1/2,r(1/2,0)}$ or $\bar\CC_{1/2,r(1/2,0)}$ representation (they have more degrees of freedom). On the other hand, suppose that $\CO_{3,i,j,\alpha\dot\alpha}$ is an IR superconformal descendant. Then it is a $\bar Q^1_{\dot\alpha}$ descendant, but this contradicts \eqref{nc}.

Next, suppose that we have 
\begin{equation}\label{nclong}
\bar Q^1_{\dot\alpha}\CO_{3,i,j,\alpha}=\lambda_a\CO_{3,i,j,\alpha\dot\alpha}\ne0~, \ \ \ (\bar Q^1)^2\CO_{3,i,j,\alpha}\ne0~.
\end{equation}
In this case, $\CO_{3,i,j,\alpha\dot\alpha}$ is highest $SU(2)_R$ weight and satisfies
\begin{equation}\label{semishort2}
(\bar Q^1)^2\CO_{3,i,j,\alpha\dot\alpha}=0~.
\end{equation}
Let us now understand where $\CO_{3,i,j,\alpha\dot\alpha}$ can sit in the IR superconformal representation theory. If $\CO_{3,i,j,\alpha\dot\alpha}$ is a superconformal primary, the shortening condition in \eqref{semishort2} is inconsistent unless $\CO_{3,i,j,\alpha\dot\alpha}$ is in a $\bar\CC_{1,r'(1/2,1/2)}$ or $\hat\CC_{1(1/2,1/2)}$ multiplet. However, \eqref{nclong} has at most the number of degrees of freedom of a long multiplet of type $\CA_{1/2,r(1/2,0)}^{\Delta}$ and therefore has fewer degrees of freedom than a $\bar\CC_{1,r'(1/2,1/2)}$ multiplet. At the same time, it has sixteen more degrees of freedom than the direct sum $\hat\CC_{1(1/2,1/2)}\oplus\bar\CB_{1/2,5/2(1/2,0)}$; however, we also require in \eqref{nclong} that $\bar Q^{1\dot\alpha}\CO_{3,i,j,\alpha\dot\alpha}\ne0$. This implies an additional multiplet with $R=3/2$, and $\CA_{1/2,r(1/2,0)}^{\Delta}$ cannot accommodate the additional degrees of freedom.

Finally, if $\CO_{3,i,j,\alpha\dot\alpha}$  is a superconformal descendant, then it must be a superconformal $\bar Q^1_{\dot\alpha}$ descendant, but this statement contradicts \eqref{nclong}. Therefore, \eqref{nclong} is inconsistent.

Next let us consider the case that 
\begin{equation}\label{nchiral}
\bar Q^1_{\dot\alpha}\CO_{3,i,j,\alpha}=0~,
\end{equation}
even after the irrelevant deformation. Then, to try to avoid a potential $\bar\CB_{1/2,r(1/2,0)}$ multiplet in the UV, let us suppose $\CO_{3,i,j,\alpha}$ is now a descendant
\begin{equation}\label{tOaadot}
\bar Q^{1\dot\alpha}\tilde\CO_{3,i,j,\alpha\dot\alpha}=\lambda_b\CO_{3,i,j,\alpha}\ \Rightarrow\ (\bar Q^1)^2\tilde\CO_{3,i,j,\alpha\dot\alpha}=0~.
\end{equation}
Let us first imagine that $\tilde\CO_{3,i,j,\alpha\dot\alpha}$ is a SUSY primary after the irrelevant deformation. Then, in the UV it satisfies the shortening condition in \eqref{tOaadot}. UV superconformal invariance implies that $\tilde\CO_{3,i,j,\alpha\dot\alpha}$ is a $\bar\CC$ (or $\hat\CC$) superconformal primary. However, \eqref{tOaadot} contradicts the spontaneous breaking of $U(1)_r$ in the flow to the IR.

Let us suppose instead that $\tilde\CO_{3,i,j,\alpha\dot\alpha}$ is a SUSY descendant. Then, it is a $\bar Q^1_{\dot\alpha}$ descendant, and we require that
\begin{equation}\label{desc}
(\bar Q^{1})^2\tilde\CO_{\alpha}=\lambda_b\CO_{3,i,j,\alpha}~.
\end{equation}
Note that in the IR SCFT, $\tilde\CO_{\alpha}$ cannot be a descendant since then it is a $\bar Q^1_{\dot\alpha}$ descendant, and \eqref{desc} cannot hold. Suppose $\tilde\CO_{\alpha}$ is an IR superconformal primary. Then, in the IR SCFT, it must be in a $\bar\CD$, $\bar\CB$, $\hat\CC$, or $\bar\CC$ multiplet with $R\ge1/2$. From \eqref{desc}, we see that $\tilde\CO_{\alpha}$ would be a member of such a multiplet with smaller scaling dimension than the operators in \eqref{BhalfgenIR}. More precisely, the irrelevant couplings have non-positive mass dimension and so in the IR SCFT $\Delta(\tilde\CO_{\alpha})\le\Delta(\CO_{3,i,j,\alpha})-1=5/2$. Then, the only possibility is that $\tilde\CO_{\alpha}$ is an IR $\bar\CD_{1/2(1/2,0)}$ primary. However, there are always fewer such operators than $\bar\CB$ operators of the type described in \eqref{BhalfgenIR} as long as the Coulomb branch is genuine (i.e., just consisting of $U(1)^N$ super-Maxwell theory at generic points). More generally, as long as the Coulomb branch is freely generated, we can repeat the analysis starting around \eqref{nc} for $\bar\CD_{1/2(1/2,0)}$ to arrive at

\bigskip
\noindent
{\bf Statement 3:} In any rank $N\ge 2$ 4d $\CN=2$ SCFT with an $N$-dimensional freely generated Coulomb branch, there is at least one $r$ such that $\bar\CB_{1/2,r(1/2,0)}$ is in the spectrum.

\bigskip
\noindent
In particular, this result implies that $\bar\CB_{1/2,r(1/2,0)}$ multiplets are ubiquitous: most known higher-rank $\CN=2$ SCFTs have freely generated Coulomb branches.

\subsec{$\bar{\CB}$ multiplets and the Witten anomaly}\label{Wanomsec}
In theories with an $Sp(n)$ global flavor symmetry (here $Sp(1)\cong SU(2)$), we may find a $\mathbb{Z}_2$-valued 't Hooft anomaly arising from large (background) gauge transformations associated with $\pi_4(Sp(N))\cong\mathbb{Z}_2$ \cite{Witten:1982fp}. We will argue that, under fairly lax assumptions, any theory possessing such an anomaly has a $\bar\CB_{1,r(0,0)}$ multiplet (here $r$ is the $U(1)_r$ charge of a generator of the Coulomb branch chiral ring).

To understand this statement, we note that these anomalies are invariants of $Sp(n)$-preserving RG flows. Since Coulomb branch operators are necessarily uncharged under flavor symmetries \cite{Buican:2013ica,Buican:2014qla}, RG flows onto the Coulomb branch triggered by turning on vevs for Coulomb branch chiral primaries preserve $Sp(n)$ flavor symmetry.

Therefore, let us assume that the theory has a Coulomb branch, and let us study flows onto this space. To get a handle on the possibilities in the IR, note that the arguments in \cite{Garcia-Etxebarria:2017crf} show the $\mathbb{Z}_2$-valued anomaly cannot be saturated by a TQFT (see \cite{Cordova:2018acb} for another application of this fact). Therefore, on the Coulomb branch, we require massless degrees of freedom that match the UV $Sp(n)$ anomaly.

A simple example is the $SU(2)$ $\CN=4$ SYM theory. This theory has, from the $\CN=2$ perspective, an $Sp(1)$ flavor symmetry under which the components of the adjoint hypermultiplet, $(Q^a,\tilde Q^a)$ transform as doublets.\footnote{The corresponding holomorphic moment maps are $Q^aQ_a$, $\tilde Q^a\tilde Q_a$, and $Q^a\tilde Q_a$.} Since $a=1,2,3$, we have an odd number of doublets and hence a $\mathbb{Z}_2$ anomaly. Now, consider the $Sp(1)$-preserving RG flow gotten by turning on a vev for the vector multiplet scalars. This vev results in an IR theory which is just $U(1)$ $\CN=4$ SYM. The abelian effective theory has a single doublet, $(q,\tilde q)$, which realizes an $Sp(1)$ symmetry and therefore also gives rise to a $\mathbb{Z}_2$ anomaly.

A more elaborate example involves the rank-one theory with $Sp(5)$ symmetry discovered in \cite{Argyres:2010py}. There we can also turn on an expectation value for the Coulomb branch generator and flow to a Coulomb branch which has five hypermultiplets at generic points. These fields realize the $Sp(5)$ symmetry and exhibit the $\mathbb{Z}_2$ anomaly as well (the hypermultiplets form a single {\bf 10} representation of $Sp(5)$).

Now, let us suppose we flow onto the Coulomb branch of a theory exhibiting the $\mathbb{Z}_2$ 't Hooft anomaly by spontaneously breaking superconformal symmetry via a vev for a $\bar\CE$ primary. Since no IR TQFT saturates the anomaly, we have a decoupled massless sector furnishing an $Sp(n)$ holomorphic moment map, $\mu$ (we assume the flavor symmetry is locally realized). Therefore, in conjunction with the Coulomb branch operator, $\phi^2$, we can construct the normal-ordered product
\begin{equation}\label{phimuW}
\CO_W:=\phi^2\mu\in\bar\CB_{1,2(0,0)}~,
\end{equation}
which is clearly a superconformal primary of the correct type (recall that $\mu$ has $R=1$ and $r=0$). Note that this multiplet transforms in the adjoint of $Sp(n)$. For technical reasons that will become apparent later, let us assume that the IR theory is completely free (i.e., it consists of free vectors and hypers).

Next, suppose we deform the theory and flow back up to the UV. What can \eqref{phimuW} come from in the UV? We repeat the logic beginning around \eqref{nc}. In particular, to avoid a $\bar\CB_{1,r(0,0)}$ multiplet in the UV, we need to have that \eqref{phimuW} becomes a SUSY descendant or is no longer chiral after turning on some (generally infinite) irrelevant couplings, $\left\{\lambda_a\right\}$, and flowing back to the UV.

To that end, first suppose $\CO_W$ is a SUSY primary satisfying\footnote{As in the discussion around \eqref{nc}, we make the same simplifying assumptions on the appearance of the $\lambda_a$ in our multiplets described below. This is for the sake of simplicity of presentation.}
\begin{equation}\label{ncW}
\bar Q^1_{\dot\alpha}\CO_{W}=\lambda_a\CO_{W\dot\alpha}\ne0~, \ \ \ (\bar Q^1)^2\CO_{W}=0~.
\end{equation}
At short distances, the superconformal primary, $\CO_{W}$, satisfies the shortening condition in \eqref{ncW} and is therefore in either a $\bar\CB_{1,r(0,0)}$ or a $\bar\CC_{1,r(0,0)}$ multiplet (note that a $\bar\CD_{1,(0,0)}$ multiplet clearly contains too few degrees of freedom relative to the IR). As in the related proof of statement 3, neither possibility is consistent.

Indeed, let us first suppose that $\CO_{W\dot\alpha}$ is an IR superconformal primary. Then it is a primary of an IR $\bar\CC_{3/2,r'(0,1/2)}$ or $\hat\CC_{3/2(0,1/2)}$ multiplet. However, such multiplets cannot sit inside a UV $\bar\CB_{1,r(0,0)}$ or $\bar\CC_{1,r(0,0)}$ representation (they have more degrees of freedom). On the other hand, suppose that $\CO_{W\dot\alpha}$ is an IR superconformal descendant. Then it is a $\bar Q^1_{\dot\alpha}$ descendant, but this contradicts \eqref{ncW}.

Next, suppose that we have 
\begin{equation}\label{nclongW}
\bar Q^1_{\dot\alpha}\CO_W=\lambda_a\CO_{W\dot\alpha}\ne0~, \ \ \ (\bar Q^1)^2\CO_{W}\ne0~.
\end{equation}
In this case, $\CO_{W\dot\alpha}$ is highest $SU(2)_R$ weight and satisfies
\begin{equation}\label{semishort2W}
(\bar Q^1)^2\CO_{W\dot\alpha}=0~.
\end{equation}
Let us now understand where $\CO_{W\dot\alpha}$ can sit in the IR superconformal representation theory. If $\CO_{W\dot\alpha}$ is a superconformal primary, the shortening condition in \eqref{semishort2W} is inconsistent unless $\CO_{W\dot\alpha}$ is in a $\bar\CC_{3/2,r'(0,1/2)}$ or $\hat\CC_{3/2(0,1/2)}$ multiplet. However, \eqref{nclongW} has at most the number of degrees of freedom of a long multiplet of type $\CA_{1,r(0,0)}^{\Delta}$ and therefore has fewer degrees of freedom than a $\bar\CC_{3/2,r'(0,1/2)}$ multiplet or a $\hat\CC_{3/2(0,1/2)}$ multiplet.

Finally, if $\CO_{W\dot\alpha}$  is a superconformal descendant, then it must be a superconformal $\bar Q^1_{\dot\alpha}$ descendant, but this statement contradicts \eqref{nclongW}. Therefore, \eqref{nclongW} is inconsistent.

Next let us consider the case that 
\begin{equation}\label{nchiralW}
\bar Q^1_{\dot\alpha}\CO_{W}=0~,
\end{equation}
even after the irrelevant deformation. Then, to try to avoid a potential $\bar\CB_{1,r(0,0)}$ multiplet in the UV, let us suppose $\CO_{W}$ is now a descendant
\begin{equation}\label{tOaadotW}
\bar Q^{1\dot\alpha}\tilde\CO_{W\dot\alpha}=\lambda_b\CO_{W}\ \Rightarrow\ (\bar Q^1)^2\tilde\CO_{W\dot\alpha}=0~.
\end{equation}
Let us first imagine that $\tilde\CO_{W\dot\alpha}$ is a SUSY primary after the irrelevant deformation. Then, in the UV it satisfies the shortening condition in \eqref{tOaadotW}. UV superconformal invariance implies that $\tilde\CO_{W\dot\alpha}$ is a $\bar\CC$ (or $\hat\CC$) superconformal primary. However, \eqref{tOaadotW} contradicts the spontaneous breaking of $U(1)_r$ in the flow to the IR.

Let us suppose instead that $\tilde\CO_{W\dot\alpha}$ is a SUSY descendant. Then, it is a $\bar Q^1_{\dot\alpha}$ descendant, and we require that
\begin{equation}\label{descW}
(\bar Q^{1})^2\tilde\CO_{W}=\lambda_b\CO_{W}~.
\end{equation}
Note that in the IR SCFT, $\tilde\CO_{W}$ cannot be a descendant since then it is a $\bar Q^1_{\dot\alpha}$ descendant, and \eqref{descW} cannot hold. Suppose $\tilde\CO_{W}$ is an IR superconformal primary. Then, in the IR SCFT, it must be in a $\bar\CD$, $\bar\CB$, $\hat\CC$, or $\bar\CC$ multiplet with $R\ge0$. From \eqref{descW}, we see that $\tilde\CO_{W}$ would be a member of such a multiplet with smaller scaling dimension than the operators in \eqref{phimuW}. More precisely, the irrelevant couplings have non-positive mass dimension and so in the IR SCFT $\Delta(\tilde\CO_{W})\le\Delta(\CO_{W})-1=3$. Then, the only possibilities are that $\tilde\CO_{W}$ is an IR $\bar\CD_{1(0,0)}$, $\bar\CB_{1/2,2(0,0)}$, $\hat\CC_{1/2(0,0)}$, or a $\bar\CC_{0,1(0,0)}$ primary. From \eqref{phimuW}, it is clear that this operator must transform in the adjoint of $Sp(n)$.

Let us suppose that we can go to a point on the Coulomb branch where the theory is completely free (this includes free hypermultiplets). Then, we see that to get an $Sp(n)$ adjoint primary, we require two hypermultiplet scalars. This logic leaves a single free vector scalar for us to adjoin to get an operator of dimension three. As a result, we can immediately rule out the $\bar\CB$, $\hat\CC$, and $\bar\CC$ options.\footnote{For the latter case, note that to get $R=0$ we should anti-symmetrize the free hyper $SU(2)_R$ indices. This implies that the operator is not in the adjoint of the flavor symmetry.} 

We are left with the $\bar\CD$ option. For higher-rank Coulomb branches, we can always build more $\bar\CB$ operators of type \eqref{phimuW}, but in the rank-one case we cannot. Instead, for rank one, we can repeat the logic around \eqref{ncW} for $\phi\mu\in\bar\CD_{1(0,0)}$ rather than $\phi^2\mu\in\bar\CB_{1,2(0,0)}$. Therefore, we arrive at

\bigskip
\noindent
{\bf Statement 4:} Consider a 4d $\CN=2$ SCFT with an $Sp(n)$ flavor symmetry having a non-vanishing $\mathbb{Z}_2$-valued anomaly. If the theory possesses points on the Coulomb branch consisting purely of free fields, then it has a $\bar\CB_{1,r(0,0)}$ multiplet, where $r$ is the $U(1)_r$ charge of a Coulomb branch generator. This multiplet transforms in the adjoint of $Sp(n)$.\footnote{We can arrive at this result more simply if we are willing to invoke the lore that a mixed branch implies the existence of operators in the \eqref{BEOPE} channel. Indeed, the existence of the mixed branch follows from the discussion around \eqref{phimuW}.}

\subsec{Free theories and a conjecture on general 4d $\CN=2$ SCFTs}
In this section, we would like to prove a few statements about the spectrum of $\bar\CB$ multiplets in free 4d $\CN=2$ SCFTs and then use these statements to conjecture a general constraint on 4d $\CN=2$ SCFTs.

\begin{table}[]
    \centering
    \begin{tabular}{c c c c c c}
        $$ & $R$ & $r$ & $j$ & $E$ & $\delta$ \\
        $\phi_i$ & 0 & 1 & 0 & 1 & 0 \\ 
        $\lambda^1_{i,\alpha}$ & 1/2 & 1/2 & $\pm$1/2 & 3/2 & 0 \\ 
        $q_a$ & 1/2 & 0 & 0 & 1 & 0 \\ 
        $\tilde{q}_a$ & 1/2 & 0 & 0 & 1 & 0 \\ 
    \end{tabular}
    \caption{List of chiral fields in a theory with $N$ abelian free vector multiplets ($i=1,\cdots,N$) and $M$ free hypermultiplets ($a=1,\cdots,M$).}
    \label{tab: Free Fields}
\end{table}

To that end, we begin with the following observation:

\bigskip
\noindent
{\bf Fact 1:} In a theory of $N$ free vectors and $M$ free hypermultiplets, all $\bar\CB_{R,r(j,0)}$ and $\bar\CD_{R(j,0)}$ multiplets satisfy $j\le R$.

\bigskip
\noindent
To understand this statement, note that a free vector and a free hypermultiplet have the chiral fields listed in table \ref{tab: Free Fields}. Let us construct highest $SU(2)_R$ and Lorentz-weight primaries of $\bar\CB$ and $\bar\CD$ multiplets in a theory of $N$ free vectors and $M$ free hypers. The only sources of $j$ are the gauginos, $\lambda^1_{i,\alpha}$ (in particular, derivatives would lead to operators that are trivial in the chiral ring and hence not highest-weight primaries of $\bar\CB$ or $\bar\CD$ multiplets). Since $j\le R$ for the gaugino (and all other chiral fields), any word constructed out of letters in table \ref{tab: Free Fields} has $j\le R$. $\square$

\bigskip
While the above argument provides a bound on spin versus $SU(2)_R$ quantum numbers, it is natural to ask if we can realize all of the above multiplets with $j\le R$. Indeed, this is the case:

\bigskip
\noindent
{\bf Fact 2:} In a theory of $N=2n$ free vectors and $M\ge1$ free hypermultiplets, there exist values of $r$ such that we have at least one $\bar\CB_{R,r(j,0)}$ and one $\bar\CD_{R(j,0)}$ multiplet for all $R\le (N-1)/2$ and $j\le R$.

\bigskip
\noindent
To derive this set of facts, note that
\begin{equation}
\CO_{u,v}=q^u\phi_1^v~,
\end{equation}
is a highest-weight primary of a $\bar\CB_{u/2,v(0,0)}$ multiplet if $v>1$ and a $\bar\CD_{u/2(0,0)}$ multiplet if $v=1$ (clearly the above component fields are superconformal primaries and therefore so too is the product).

Next, let us construct $\bar\CD_{(2m-1)/2((2m-1)/2,0)}$ and $\bar\CD_{m-1(m-1,0)}$ multiplets. To that end, note that
\begin{eqnarray}
\CO_{(1,2m),+^{2m-1}}&:=&\sum_{i=1}^{2m}(-1)^i\phi_i\prod_{j=1,j\ne i}^{2m}\lambda^1_{j,+}~, \ \ \ 1\le m\le n~,\cr
\CO_{(1,2m-1),+^{2m-2}}&:=&\sum_{i=1}^{2m-1}(-1)^i\phi_i\prod_{j=1,j\ne i}^{2m-1}\lambda^1_{j,+}~, \ \ \ 1<m\le n~,
\end{eqnarray}
are highest-weight primaries of a $\bar\CD_{(2m-1)/2,((2m-1)/2,0)}$ multiplet and a $\bar\CD_{m-1,(m-1,0)}$ multiplet respectively. Indeed, $r=1+j$ by construction, and the above operators are annihilated by all $S$ and $\bar S$ supercharges. Now consider the operators
\begin{equation}
q^u\CO_{(1,2m),+^{2m-1}}\in\bar\CD_{(2m-1+u)/2((2m-1)/2,0)}~,\ \ \ q^u\CO_{(1,2m-1),+^{2m-2}}\in\bar\CD_{m-1+u/2(m-1,0)}~,
\end{equation}
This operator is a primary of a $\bar\CD$ multiplet since $q$ is a superconformal primary with $r=j=0$. Allowing $u$ to be arbitrary, we have proven our claim for $\bar\CD$ multiplets.

To arrive at the claim for $\bar\CB$ multiplets, we can simply take the above operators and multiply by $\phi_i$
\begin{eqnarray}
\phi_iq^u\CO_{(1,2m),+^{2m-1}}&\in&\bar\CB_{(2m-1+u)/2,(2m+3)/2((2m-1)/2,0)}~,\cr \phi_iq^u\CO_{(1,2m-1),+^{2m-2}}&\in&\bar\CB_{m-1+u/2,m+1(m-1,0)}~,
\end{eqnarray}
Indeed, this statement follows from the fact that $\phi_i$ is a superconformal primary and has $r=1$ and $j=0$. $\square$

\bigskip
This logic also implies the following fact:

\bigskip
\noindent
{\bf Fact 3:} For any $j\le R$, there exist values of $r$ and a local unitary 4d $\CN=2$ SCFT, $\CT$, such that $\bar\CB_{R,r(j,0)}$ and $\bar\CD_{R(j,0)}$ is in the spectrum of $\CT$.

\bigskip
If we consider the full set of free theories, it is reasonable to imagine that we see all possible superconformal representations up to deformations of the $U(1)_r$ charge. The heuristic reason for this belief is that, in a general theory, we expect interactions to lead to new null states. At the same time, interactions cannot change the $SU(2)_R$ and Lorentz-spin quantization (they can only lead to changes in the quantization of $U(1)_r$).

Moreover, the general arguments of \cite{Manenti:2019kbl} show that any local 4d $\CN=2$ SCFT only has $\bar\CD_{R(j,0)}$ multiplets for $j\le R$. Therefore, we are led to the following conjecture:

\bigskip
\noindent
{\bf Conjecture:} $\bar\CB_{R,r(j,0)}$ multiplets with $j>R$ are forbidden in general local unitary 4d $\CN=2$ SCFTs.\footnote{In fact, the analysis of \cite{Manenti:2019kbl} can be used to directly prove our conjecture for the special case of $r<j+2$. We thank A.~Manenti for pointing this fact out to us and for collaboration on upcoming work on this conjecture \cite{toApp}.}

\bigskip
\noindent
Note that a bound of the above form on $\bar\CB$ implies the bound on $\bar\CD$ found in \cite{Manenti:2019kbl}. Indeed, suppose this were not the case. Then, we would have a $\bar\CD_{R(j,0)}$ multiplet with $j>R$. Taking the product of the corresponding highest-weight primary with a free vector $\phi$ primary would give a $\bar\CB$ multiplet with $j>R$. On the other hand, note that the $\bar\CD$ bound does not imply the $\bar\CB$ bound since not all $\bar\CB$ operators need to come from a product of the form $\bar\CD\times\bar\CE$.

Note also that the above bound on $\bar\CB$ implies a known bound on $\bar\CE_{r(j,0)}$ ruling out $j>0$ in these latter multiplets \cite{Manenti:2019jds}. Indeed, suppose that there were a $j>0$ such that $\bar\CE_{r(j,0)}$ existed in some SCFT, $\CT$. Then, we could take $N$ decoupled copies of $\CT$ to get $\bar\CE_{r'(j',0)}$ with arbitrarily large $j'>0$. Therefore, multiplying with a free hypermultiplet would give $\bar\CB$ violating the conjecture by an arbitrarily large amount.

\newsec{$\bar{\CB}$ multiplets in Rank-one theories}
We have shown that, from relatively minimal assumptions, theories with $\CN>2$ SUSY, higher rank and a conformal manifold, a $\mathbb{Z}_2$-valued $Sp(N)$ anomaly, or with freely generated higher-dimensional Coulomb branches must possess $\bar\CB$ multiplets. Therefore, in this section, we specialize to rank-one theories that do not satisfy any of these properties (and also have no mixed branches) in order to understand whether any of these theories have $\bar\CB$ multiplets.

We focus on a subset of such theories that can be described by certain non-conformal $\CN=1$ Lagrangians with accidental IR enhancement to $\CN=2$ (e.g., see \cite{Maruyoshi:2016aim}). In this context, a necessary condition for a Lagrangian to be useful in carrying out precision spectroscopy is for the IR superconformal $U(1)_r$ and $SU(2)_R$ Cartan to be visible in the UV and unbroken along the RG flow. This property is typically absent in $\CN=2$ RG flows and hence explains the utility of constructions involving accidental SUSY enhancement.

\subsection{General Strategy}\label{Gen}
As described in section \ref{Bbargen}, the spectrum of all multiplets in the FCS except the $\bar\CB$ multiplets can easily be determined either from a Seiberg-Witten description (in the case of the $\bar\CE$ multiplets) or from the associated 2d VOA (in the case of the $\hat\CB$ and $\bar\CD$ multiplets). Therefore, to get a handle on the $\bar\CB$ spectrum, our strategy will be similar to the one adopted in \cite{Bhargava:2022cuf}: we will study rank-one $\CN=2$ theories with weakly coupled UV descriptions in terms of an $\CN=1$ gauge theory that is connected to our SCFT of interest by a sufficiently \lq\lq smooth" RG flow (here we require that the superconformal IR symmetry is visible along the RG flow).

More precisely, we will use the fields in these Lagrangian descriptions to write down the full set of gauge-invariant local operators that are candidates to generate the $\CN=1$ chiral ring (we assume the RG flow only acts to truncate this ring). This is the set of chiral operators that cannot be expressed as a product of two nontrivial chiral operators.\footnote{The identity operator is defined as the trivial chiral operator.} We whittle down this list by demanding that these generators sit in certain IR $\CN=2$ superconformal representations.

In fact, in all the rank-one theories we will consider, we never find a situation where an operator in this smaller list belongs to a $\bar{\CB}$ multiplet. While we do not have a full understanding of when $\bar\CB$ multiplets do and do not furnish generators of the chiral ring more generally, this observation plays an important role in further results we derive about the spectrum of chiral operators in the particular theories we study.

As a consistency check of our method, we can make contact with known results on the non-$\bar\CB$ part of the FCS when characterizing our generators. We can then use these operators to exhaustively attempt to construct $\bar\CB$ operators through chiral ring fusion.

In particular, rank-one theories have a one-complex-dimensional Coulomb branch. We will study examples where the corresponding Coulomb branch chiral ring is freely generated. At the level of superconformal representation theory, this means that we have a single $\bar\CE_r$ generator giving rise to an $\CN=2$ chiral ring of primaries via the $n$-fold OPEs (for $n\in\mathbb{Z}_{>0}$), $\bar\CE_{r}^{\times n}\ni\bar\CE_{rn}$.

Most of the rank-one theories we study also have an $\CN=2$ flavor symmetry. As we have seen in previous sections, the Noether currents sit in corresponding $\hat\CB_1$ multiplets transforming in the adjoint of the flavor symmetry. These multiplets are associated with the Higgs branch. For all the theories we consider, the $\hat\CB_1$ multiplets generate the Higgs branch chiral ring.

Moreover, all the examples we study here can be mapped to known 2d VOAs, and we can use these associated VOAs to conclude that there are no $\bar\CD$ multiplets in our spectra. As a result, the $\bar\CB$ generation channel in \eqref{DDOPE} is not available (recall that none of the theories we study here have $\CN>2$ SUSY and so the result of section \ref{N3susy} does not apply).

Therefore, any $\bar{\CB}$ multiplets in our theories of interest must be linear combinations of normal-ordered products of chiral operators sitting in $\bar\CE_r$ and / or $\hat\CB_1$ multiplets. We construct all such products allowed by the OPE constraints and chiral ring relations in the theory, and we consider their linear combinations. 

Two kinds of $\bar{\CB}$ multiplets will be of particular interest in our analysis, so we single them out beforehand. These are obtained from the OPE in \eqref{BEOPE} and the second OPE in \eqref{EEdesc}. In the rank-one theories we consider, these $\bar\CB$ multiplets often vanish as a result of chiral ring relations arising from the dynamics of the $\CN=1\to\CN=2$ RG flows.
 
We now proceed to our analysis of the $\CN=1$ chiral spectra of individual rank-one SCFTs. For simplicity, we stick to isolated theories of Argyres-Douglas type and to the $SU(2)$ theory with four flavors.
\subsection{The $(A_1,A_2)\cong{\rm MAD}\cong H_0\cong \mathfrak{a}_0$ theory}
The chiral ring of this theory was analyzed in detail in \cite{Bhargava:2022cuf}, and so we merely summarize the story here. Recall that this is the original Argyres-Douglas theory \cite{Argyres:1995jj} (referred to in \cite{Buican:2021elx,Bhargava:2022cuf,Bhargava:2022yik} as the \lq\lq Minimal" Argyres-Douglas (MAD) theory). It has a Coulomb branch chiral ring generator of dimension $6/5$ sitting as a primary in a $\bar\CE_{6/5}$ multiplet. The theory has no Higgs branch, and, consistent with this fact, the associated 2d VOA is the Lee-Yang Virasoro vacuum module \cite{Cordova:2015nma}. Arguments in \cite{Buican:2021elx} then imply that there are no $\hat\CB$ or $\bar\CD$ multiplets in this theory.

\begin{table}[]
    \centering
    \begin{tabular}{c c c c c c c c c}
         $\text{Fields}$ & $SU(2)_{\text{gauge}}$ & $R$ & $r$\\
         $\phi$ & $\text{adj}$ & 0 & 1/5 \\ 
         $\lambda_{\alpha}$ & $\text{adj}$ & 1/2 & 1/2  \\ 
         $q$ & 2 & 1/2 & 2/5 \\ 
         $\tilde{q}$ & 2 & 1/2 & -1/5 \\ 
         $M$ & 1 & 0 & 6/5 \\ 
         $X$ & 1 & 1 & 3/5 \\ 
    \end{tabular}
    \caption{UV fields in the $\CN=1$ description of the $(A_1,A_2)$ theory \cite{Maruyoshi:2016tqk,Benvenuti:2018bav}. Here $r$ and $R$ are the IR $U(1)_r$ and $SU(2)_R$ Cartan respectively.}
    \label{tab:A2 Fields}
\end{table}

The $\CN=1$ $SU(2)$ gauge theory Lagrangian with fields in Table \ref{tab:A2 Fields}\footnote{Our naming conventions differ slightly from \cite{Bhargava:2022cuf}. In particular, $\tilde q\to q'$.} and superpotential \cite{Maruyoshi:2016tqk,Benvenuti:2018bav}
\begin{equation}\label{MADW}
    W=X\phi^2+M\phi q'q'+\phi qq~,
\end{equation}
was used in \cite{Bhargava:2022cuf} to argue that there are no $\bar\CB$ multiplets and 
\begin{equation}
{\rm FCS}_{(A_1,A_2)}=\langle\CE_{6/5}\rangle~,
\end{equation}
where the chiral operators in $\bar\CE_{6/5}$ are located as follows
\bea \label{ElocMAD}
M\in (0,0)_{0,6/5} &\xrightarrow{Q^1_\alpha}& \phi\lambda_{\alpha}\in \left({1\over2},0\right)_{1/2,7/10}\xrightarrow{(Q^1)^2 }  qq'\in(0,0)_{1,1/5}~.
\eea

The basic idea of the proof in \cite{Bhargava:2022cuf} was to write down all possible chiral ring generators and argue that none can sit in $\bar\CB$ representations (i.e., generators cannot sit at locations described in \eqref{Chiralloc2}). Then, dynamical constraints from the superpotential \eqref{MADW} rule out chiral ring products of operators in \eqref{ElocMAD} giving rise to a $\bar\CB$ multiplet (crucially, the channel described in \eqref{EEdesc} does not create $\bar\CB$ chiral operators).

\subsection{The $(A_1,A_3)\cong H_1\cong \mathfrak{a}_1$ theory}
This theory was not analyzed in \cite{Bhargava:2022cuf}. It is the simplest Argyres-Douglas theory with flavor symmetry \cite{Argyres:1995xn} ($SO(3)$ for the untwisted Hilbert space in this case \cite{Buican:2021xhs,Closset:2021lhd}). The Coulomb branch chiral ring generator has dimension $4/3$ and sits as a primary in an $\bar\CE_{4/3}$ multiplet. Unlike the previous case, the theory has a one-quaternionic-dimensional Higgs branch with the Higgs branch chiral ring generated by the holomorphic $SO(3)$ moment map $\mu^a\in\hat\CB_1^a$ (here $a=\pm,0$ indicates $SO(3)$ flavor weight) subject to the relation
\begin{equation}\label{HiggsRelA1A3}
\mu^+\mu^-\sim(\mu^0)^2~.
\end{equation}

This theory also has a known associated VOA \cite{Buican:2015ina}: the $\widehat{su(2)}_{-4/3}$ affine Kac-Moody (AKM) algebra. This fact allows us to immediately rule out $\bar\CD$ multiplets. Indeed, $\widehat{su(2)}_{-4/3}$ is (strongly) generated by the affine current (related, by the map in \cite{Beem:2013sza}, to $\mu^a$ which has $r=0$). This means that any operator in the 2d VOA is built from normal-ordered products of (derivatives) of this current. Since the procedure in \cite{Beem:2013sza} used to construct the VOA respects the $U(1)_r$ symmetry, we conclude that all Schur operators in the $(A_1, A_3)$ theory are $U(1)_r$ neutral. Since $\bar\CD$ Schur operators necessarily have $r\ne0$, they cannot be present.

\begin{table}[]
    \centering
    \begin{tabular}{c c c c c}
         $\text{Fields}$ & $SU(2)_{\text{gauge}}$ & $R$ & $r$  \\
         $\phi$ & $\text{adj}$ & 0 & 1/3 \\ 
         $\lambda_{\alpha}$ & $\text{adj}$ & 1/2 & 1/2 \\ 
         $q$ & 2 & 1/2 & -1/6 \\ 
         $\tilde{q}$ & 2 & 1/2 & -1/6 \\ 
         $\alpha_0$ & 1 & 0 & 4/3 \\ 
         $\beta_2$ & 1 & 1 & 1/3 \\ 
    \end{tabular}
    \caption{UV fields in the $\CN=1$ description of the $(A_1,A_3)$ theory \cite{Maruyoshi:2016tqk,Benvenuti:2017lle}. Here $r$ and $R$ are the IR $U(1)_r$ and $SU(2)_R$ Cartan.}
    \label{tab:A3 fields}
\end{table}

Therefore, the FCS can at most consist of $\bar\CE$, $\hat\CB$, and $\bar\CB$ multiplets. To get a handle on these latter multiplets, we consider the $\CN=1$ Lagrangian with fields given in Table \ref{tab:A3 fields} and superpotential \cite{Maruyoshi:2016tqk,Benvenuti:2017lle}
\begin{equation}\label{WA1A3}
W=\alpha_0 q\tilde{q}+\beta_2\phi^2~.
\end{equation}
As in the $(A_1, A_2)$ case, the UV theory is an $SU(2)$ $\CN=2$ gauge theory. However, this time the superpotential preserves the $SU(2)$ flavor symmetry under which $(q,\tilde q)$ transforms as a doublet. This fact is crucial in order to reproduce the IR symmetry discussion around \eqref{HiggsRelA1A3}.

\begin{table}[]
    \centering
    \begin{tabular}{ c c c c c c}
         $\text{Operator}$ & $R$ & $r$  & $j$  \\ 
         $\alpha_0$ & 0 & 4/3 & 0  \\ 
         $\beta_2$ & 1 & 1/3 &  0 \\ 
         $\phi^2$ & 0 & 2/3 &  0  \\ 
         $\lambda^2$ & 1 & 1 & 0 \\ 
         $q\tilde{q}$ & 1 & -1/3 & 0 \\ 
         $\phi qq$ & 1 & 0 & 0  \\ 
         $\phi q\tilde{q}$ & 1 & 0 & 0 \\ 
         $\phi\tilde{q}\tilde{q}$ & 1 & 0 & 0 \\ 
         $\phi\lambda_{\alpha}$ & 1/2 & 5/6 & 1/2 \\ 
         $qq\lambda_{\alpha}$ & 3/2 & 1/6 & 1/2 \\ 
         $q\tilde{q}\lambda_{\alpha}$ & 3/2 & 1/6 & 1/2 \\ 
         $\tilde{q}\tilde{q}\lambda_{\alpha}$ & 3/2 & 1/6 & 1/2 \\ 
         $\phi qq\lambda_{\alpha}$ & 3/2 & 1/2 & 1/2 \\ 
         $\phi q\tilde{q}\lambda_{\alpha}$ & 3/2 & 1/2 & 1/2 \\ 
         $\phi \tilde{q}\tilde{q}\lambda_{\alpha}$ & 3/2 & 1/2 & 1/2 \\ 
         $\phi\lambda_{\alpha}\lambda_{\beta}$ & 1 & 4/3 & 1 \\ 
         $qq\lambda_{\alpha}\lambda_{\beta}$ & 2 & 2/3 & 1 \\ 
         $q\tilde{q}\lambda_{\alpha}\lambda_{\beta}$ & 2 & 2/3 & 1 \\ 
         $\tilde{q}\tilde{q}\lambda_{\alpha}\lambda_{\beta}$ & 2 & 2/3 & 1 \\ 
         $\lambda_{\alpha}\lambda_{\beta}\lambda_{\gamma}$ & 3/2 & 3/2 & 3/2 \\ 
    \end{tabular}
    \caption{List of $\CN=1$ chiral generator candidates in the $(A_1,A_3)$ theory. Here $r$ and $R$ are the IR $U(1)_r$ and $SU(2)_R$ Cartan, and $j$ is the left spin.}    \label{tab:A3 generators}
\end{table}

Let us first study the potential chiral ring generators. To that end, we have listed the candidates in Table \ref{tab:A3 generators}. In compiling this list, we have used the fact that $\delta^{ab}\sim{\rm Tr}(T^aT^b)$ and $\epsilon^{abc}\sim{\rm Tr}([T^a,T^b]T^c)$ to express any chiral ring generator in terms of at most three adjoints. Note that this does not imply that the chiral ring generators will obey classical relations in the quantum theory.\footnote{Indeed, as in footnote 15 of \cite{Bhargava:2022cuf}, we can argue that generators in the quantum theory are built from traces involving two or three adjoints.\label{quantum}}

We begin by identifying the known $\bar\CE$ and $\hat\CB$ generators of the FCS in terms of the operators in Table \ref{tab:A3 generators}. To that end, note that unitarity bounds imply that the chiral operators in the $\hat\CB_1$ multiplets and the primary and level-one descendants of $\bar\CE_{4/3}$ cannot be composites built out of products of gauge-invariant operators (the level-two descendant of $\bar\CE_{4/3}$ can at most be built out of a product of two gauge invariant operators). Therefore, we can immediately identify
\begin{equation}\label{BhatlocA1A3}
\mu^+=\phi qq\in \hat\CB_1^+~, \ \ \ \mu^0=\phi q\tilde q\in \hat\CB_1^0~, \ \ \ \mu^-=\phi \tilde q\tilde q\in \hat\CB_1^-~.
\end{equation}
For the $\bar\CE_{4/3}$ multiplet we have
\bea \label{ElocA1A3}
\alpha_0\in (0,0)_{0,4/3} &\xrightarrow{Q^1_\alpha}& \phi\lambda_{\alpha}\in \left({1\over2},0\right)_{1/2,5/6}\xrightarrow{(Q^1)^2 }  \beta_2\in(0,0)_{1,1/3}~.
\eea
We have mapped $\beta_2$ to the level-two descendant of $\bar\CE_{4/3}$ using the fact that there is no other candidate built from a single generator in Table \ref{tab:A3 generators} or a product of two such generators that has the correct quantum numbers (by construction, $\phi^2$ decouples from the IR chiral ring).

What about $\bar\CB$ generators? Our analysis so far implies that any chiral operator, $\CO$, that is not in an $\bar{\CE}$ or $\hat{\CB}$ multiplet must be in a $\bar{\CB}$ multiplet. Therefore, $\CN=2$ superconformal representation theory implies that $\CO$ must have $r>j$ (and $r>1+j$ to be a superconformal primary in such a multiplet; see \eqref{Chiralloc2}).

We only have two fields in the list of candidate chiral generators that could potential sit as $\bar\CB$ descendants, namely $\lambda^2$ and $\phi\lambda_{\alpha}\lambda_{\beta}$. Moreover, there are no candidates for $\bar{\CB}$ primaries among the list in Table \ref{tab:A3 generators}.

Now, if $\lambda^2$ is a level-one descendant, then the primary has $R=1/2$, $r=3/2$, $j=1/2$. But this is the superconformal primary of a $\bar{\CD}_{1/2(1/2,0)}$ multiplet, which we know is absent. If $\lambda^2$ is a level 2 descendant, then the primary has $R=0$, $r=2$, $j=0$, which is the superconformal primary of a $\bar{\CE}_2$ multiplet. We know that such a multiplet does not exist in the $(A_1,A_3)$ theory. Therefore, $\lambda^2$ must be trivial in the IR chiral ring.

Let us perform the same analysis for $\phi\lambda_{\alpha}\lambda_{\beta}$. If this is a level-two descendant, the primary has $R=0$ and $r=7/3$, which corresponds to an $\bar{\CE}$ multiplet. However, no such multiplet exists in the theory. If this is a level-one descendant, then the primary has $R=1/2$, $r=11/6$, $j=1/2$ (it cannot have $j=3/2$, since we would then require that $r>5/2$). However, there is no product of generators in Table \ref{tab:A3 generators} that has these quantum numbers. Therefore, $\phi\lambda_{\alpha}\lambda_{\beta}$ must be trivial in the IR FCS.

As a result, we see that we have the following characterization of the generators of the FCS and the FCS itself
\begin{equation}
\text{FCS}_{(A_1, A_3)}=\langle\bar{\CE}_{4/3}~,\ \hat{\CB}_1^a\rangle/I~,
\end{equation}
where $I$ is the ideal generated by the constraint in \eqref{HiggsRelA1A3}. We would like to understand if these generators can give rise to a $\bar\CB$ multiplet
\begin{equation}
\text{FCS}_{(A_1, A_3)}\stackrel{?}{\ni}\bar\CB~.
\end{equation}

In particular, we can try to form $\bar{\CB}$ multiplets by taking products of operators in \eqref{BhatlocA1A3} and \eqref{ElocA1A3}. At the quadratic level, there are several possibilities, that we now proceed to study.
\begin{itemize}
\item $\alpha_0^2$: This is the superconformal primary of the $\bar{\CE}_{8/3}$ multiplet. In fact, since the Coulomb branch is freely generated, $\alpha_0^k$ will be the superconformal primary of the $\bar{\CE}_{4k/3}$ multiplet for all $k\in\mathbb{N}$.
\item $\alpha_0\phi\lambda_{\alpha}$: This is the level one descendant of the $\bar{\CE}_{8/3}$ multiplet. Similar to the case above, $\alpha_0^{k-1}\phi\lambda_{\alpha}$ will be the level-one descendant of the $\bar{\CE}_{4k/3}$ multiplet for all $k\in\mathbb{N}$. 
\item $\beta_2\phi\lambda_{\alpha}$: This is ruled out by the superpotential constraint,
\begin{equation}\label{Wcons}
\frac{\partial W}{\partial\phi^a}\cdot\lambda_{\alpha}^a=0~.
\end{equation}
This constraint cannot receive quantum corrections because they would involve terms with $R=3/2$ and $r=7/6$ (recall that $\phi^2$ hits a unitarity bound and decouples).
\item $\alpha_0\beta_2$ and $(\phi\lambda_{\alpha})^2$: One linear combination of these two operators will be the level-two descendant of the $\bar{\CE}_{8/3}$ multiplet. Can we find a linear combination of these two operators that would be a superconformal primary? We see that the only candidate for the level-one descendant of such a multiplet is $\beta_2\phi\lambda_{\alpha}$. However, this has already been set to zero by \eqref{Wcons}. Therefore, no linear combination of these two operators can be a superconformal primary.
\item $\beta_2^2$ ($R=2$, $r=2/3$, $j=0$): Since $r<1+j$, this is not a superconformal primary. If it is a level-one descendant, then the primary has ($R=3/2$, $r=7/6$, $j=1/2$). The candidate operators with $R=3/2$ and $j=1/2$ are $\alpha_0^m\beta_2\phi\lambda_{\alpha}=0$ (by \eqref{Wcons}), $\alpha_0^m(\phi qq)\phi\lambda_{\alpha}$, $\alpha_0^m(\phi q\tilde{q})\phi\lambda_{\alpha}$, $\alpha_0^m(\phi \tilde q\tilde{q})\phi\lambda_{\alpha}$, but none of these can lead to $r=7/6$. If it is a level-two descendant, then the primary has ($R=1$, $r=5/3$, $j=0$). The candidate operators with $R=1$ and $j=0$ are $(\phi\lambda_{\alpha})^2$ and $\alpha_0\beta_2$, but since the level-one descendant of this potential $\bar{\CB}$ multiplet vanishes, this is ruled out. Therefore, the operator $\beta_2$ must be nilpotent in the IR chiral ring (as implied by the discussion in \cite{Komargodski:2020ved}).
\item $\alpha_0\phi qq$, $\alpha_0\phi q\tilde{q}$, and $\alpha_0\tilde q\tilde q$: These are ruled out by the superpotential constraints,
\begin{eqnarray}\label{a0cons}
\frac{\partial W}{\partial\tilde{q}^j}&=&\alpha_0q_j=0~,\cr\frac{\partial W}{\partial{q}^j}&=&\alpha_0\tilde q_j=0~.
\end{eqnarray}
Therefore,\footnote{Alternatively, we could arrive at the same conclusion by using the first equation in \eqref{a0cons} and invoking $SU(2)$ flavor covariance.}
\begin{equation}
\alpha_0qq=\alpha_0q\tilde q=\alpha_0\tilde q\tilde q=0~.
\end{equation}
These constraints cannot receive quantum corrections because they would involve operators with $R=1$ and $r=4/3$ (and having the same $SO(3)$ weight as the operators in question).
\item $(\phi\lambda_{\alpha})(\phi qq)$, $(\phi\lambda_{\alpha})(\phi q\tilde{q})$, and $(\phi\lambda_{\alpha})(\phi \tilde q\tilde{q})$ ($R=3/2$, $r=5/6$, $j=1/2$): Since $r<1+j$, this is not a superconformal primary. If it is a level-one descendant then the primary has ($R=1$, $r=4/3$, $j=0$). The candidate operators with $R=1$ and $j=0$ are $\alpha_0^m\phi qq$, $\alpha_0^m\phi q\tilde{q}$, $\alpha_0^m\phi \tilde q\tilde{q}$, $\alpha_0^m(\phi\lambda_{\alpha})^2$, and $\alpha_0^m\beta_2$. Among these operators only $\alpha_0\phi q\tilde{q}$ has compatible $r$ charge. However, it is removed by the superpotential constraint in \eqref{a0cons}. If it is a level-two descendant then the primary has $R=1/2$, $r=11/6$, $j=1/2$. The only candidate with $R=1/2 $ is $\alpha_0^m\phi\lambda_{\alpha}$, but this does not have $r=11/6$. Therefore, this operator must be trivial in the IR chiral ring.
\item $\beta_2\phi qq$, $\beta_2\phi q\tilde{q}$, and $\beta_2\phi \tilde q\tilde{q}$: These operators vanish in the chiral ring as a result of the superpotential constraints,
\begin{eqnarray}
\frac{\partial W}{\partial\phi^a}\cdot (qq)^a=\frac{\partial W}{\partial\phi^a}\cdot (q\tilde{q})^a=\frac{\partial W}{\partial\phi^a}\cdot (\tilde q\tilde{q})^a=0~.
\end{eqnarray}
These constraints cannot receive quantum corrections because they would involve other operators with $R=2$ and $r=1/3$ (and having the same $SO(3)$ weight as the operators in question) since $\phi^2$ decouples.
\end{itemize}
We see that the only products that survive at the quadratic level are $\alpha_0^2$, $\alpha_0\phi\lambda_{\alpha}$, $(\phi\lambda_{\alpha})^2$, and $\alpha_0\beta_2$. Of these, the last two cannot be superconformal primaries (nor can any of their linear combinations be). 

We therefore see that the most general product of operators from the generating set that we can write down and which is a superconformal primary has the form
\begin{equation}
\alpha_0^{m_1}~,\ \ \ (\phi qq)^{m_2}(\phi q\tilde{q})^{m_3}(\phi \tilde q\tilde{q})^{m_4}~, \ \ \ m_1~,\ m_2~,\ m_4\in\mathbb{N}~, \ \ \ m_3=0,1~.
\end{equation}
These are the Coulomb and Higgs branch operators respectively (recall the constraint in \eqref{HiggsRelA1A3} that constraints $m_3$). Therefore, there are no $\bar{\CB}$ multiplets in the $(A_1,A_3)$ theory.

\subsection{The $(A_1,D_4)\cong H_2\cong \mathfrak{a}_2$ theory}
This Argyres-Douglas theory has $SU(3)$ flavor symmetry and was originally discovered in \cite{Argyres:1995xn}. Its Coulomb branch chiral ring generator has dimension $3/2$ and is a primary in a $\bar\CE_{3/2}$ multiplet. This theory has a two-quaternionic-dimensional Higgs branch and a corresponding chiral ring generated by the holomorphic $SU(3)$ moment map transforming in the {\bf8} (adjoint) representation, $\mu^a\in\hat\CB_1^a$ (here we will take $a$ to be an adjoint index) subject to the Joseph ideal constraint.

As in the previous cases, this theory has a known associated 2d VOA \cite{Buican:2015ina}: the $\widehat{su(3)}_{-3/2}$ AKM algebra. Using the same logic we used in the case of the $(A_1, A_3)$ theory in previous subsection, we can again rule out $\bar\CD$ multplets here too.

\begin{table}[]
\centering
    \begin{tabular}{ c c c c c c c c}        
         $\text{Field}$ & $\text{Rep}$ & $U(1)_B$ & $U(1)_m$ & $R$ & $r$ \\
         $\phi$ & $\text{adj}$ & 0 & 0 & 0 & 1/2 \\
         $\lambda_{\alpha}$ & $\text{adj}$ & 0 & 0 & 1/2 & 1/2 \\         
         $q_1$ & $\Box$ & 1 & 3 & 1/2 & 1/4 \\
         $\tilde{q}_1$ & $\bar{\Box}$  & -1 & -3 & 1/2 & 1/4 \\
         $q_2$ & $\Box$ & 1 & -1 & 1/2 & -1/4 \\
         $\tilde{q}_2$ & $\bar{\Box}$ & -1 & 1 & 1/2 & -1/4 \\
         $M_2$ & 1 & 0 & 0 & 0 & $\frac{3}{2}$ \\
         $\beta_2$ & 1 & 0 & 0 & 1 & 0 \\
         \end{tabular}
\caption{UV fields in the $\CN=1$ description of the $(A_1,D_4)$ theory \cite{Agarwal:2016pjo,Song:2021dhu}. Here $r$ and $R$ are the IR $U(1)_r$ and $SU(2)_R$ Cartan. $U(1)_B$ and $U(1)_m$ are flavor symmetries corresponding to Cartans of the IR $SU(3)$ flavor symmetry. }
\label{tab:D4 fields}
\end{table}

As a result, the FCS can again at most consist of $\bar\CE$, $\hat\CB$, and $\bar\CB$ multiplets. To understand the spectrum of the $\bar\CB$ multiplets we study an $\CN=1$ Lagrangian theory with fields given in Table \ref{tab:D4 fields} and superpotential \cite{Song:2021dhu}
\begin{equation}\label{WA1D4}
    W=M_2q_2\tilde{q}_2+\phi q_1\tilde{q}_1+\beta_2\phi^2~.
\end{equation}
Note that (as in the closely related $(A_1, A_3)$ case) the $\phi^2$ operator hits a unitarity bound and decouples in the IR. Moreover, only a $SU(2)\times U(1)\subset SU(3)$ flavor symmetry is manifest. Under this symmetry, $(q_2,\tilde q_2)$ transforms as a doublet (and $q_1$, $\tilde q_1$ are singlets).

\begin{table}[]
    \centering
    \begin{tabular}{c c c c c c c c }
     Operators & $R$ & $r$ & $j$ & $U(1)_B$ & $U(1)_m$ \\
     $M_2$ & 0 & 3/2 & 0 & 0 & 0 \\ 
         $\beta_2$ & 1 & 0 & 0 & 0 & 0 \\ 
         $\phi^2$ & 0 & 1 & 0 & 0 & 0 \\ 
         $\lambda^2$ & 1 & 1 & 0 & 0 & 0 \\ 
         $\tilde{q}_1q_1$ & 1 & 1/2 & 0 & 0 & 0 \\ 
         $\tilde{q}_2q_2$ & 1 & -1/2 & 0 & 0 & 0 \\
         $\tilde{q}_2q_1$ & 1 & 0 & 0 & 0 & 4 \\ 
         $q_1q_2$ & 1 & 0 & 0 & 2 & 2 \\ 
         $\tilde{q}_1\tilde{q}_2$ & 1 & 0 & 0 & -2 & -2  \\ 
         $\tilde{q}_1q_2$ & 1 & 0 & 0 & 0 & -4  \\ 
         $\phi q_1q_1$ & 1 & 1 & 0 & 2 & 6 \\ 
         $\phi\tilde{q}_1q_1$ & 1 & 1 & 0 & 0 & 0  \\ 
         $\phi\tilde{q}_1\tilde{q}_1$ & 1 & 1 & 0 & -2 & -6 \\ 
         $\phi\tilde{q}_2\tilde{q}_2$ & 1 & 0 & 0 & -2 & 2 \\ 
         $\phi\tilde{q}_2q_2$ & 1 & 0 & 0 & 0 & 0 \\ 
         $\phi q_2q_2$ & 1 & 0 & 0 & 2 & -2 \\ 
         $\phi\tilde{q}_2q_1$ & 1 & 1/2 & 0 & 0 & 4  \\ 
         $\phi q_1q_2$ & 1 & 1/2 & 0 & 2 & 2 \\ 
         $\phi\tilde{q}_1\tilde{q}_2$ & 1 & 1/2 & 0 & -2 & -2  \\ 
         $\phi\tilde{q}_1q_2$ & 1 & 1/2 & 0 & 0 & -4 \\ 
         $\phi\lambda_{\alpha}$ & 1/2 & 1 & 1/2 & 0 & 0  \\ 
         $q_1q_1\lambda_{\alpha}$ & 3/2 & 1 & 1/2 & 2 & 6  \\ 
         $\tilde{q}_1q_1\lambda_{\alpha}$ & 3/2 & 1 & 1/2 & 0 & 0  \\ 
         $\tilde{q}_1\tilde{q}_1\lambda_{\alpha}$ & 3/2 & 1 & 1/2 & -2 & -6  \\ 
         $q_2q_2\lambda_{\alpha}$ & 3/2 & 0 & 1/2 & 2 & -2  \\ 
         $\tilde{q}_2q_2\lambda_{\alpha}$ & 3/2 & 0 & 1/2 & 0 & 0 \\ 
         $\tilde{q}_2\tilde{q}_2\lambda_{\alpha}$ & 3/2 & 0 & 1/2 & -2 & 2  \\ 
         $\tilde{q}_2q_1\lambda_{\alpha}$ & 3/2 & 1/2 & 1/2 & 0 & 4  \\ 
         $q_1q_2\lambda_{\alpha}$ & 3/2 & 1/2 & 1/2 & 2 & 2  \\ 
         $\tilde{q}_1\tilde{q}_2\lambda_{\alpha}$ & 3/2 & 1/2 & 1/2 &-2 & -2  \\ 
         $\tilde{q}_1q_2\lambda_{\alpha}$ & 3/2 & 1/2 & 1/2 &0 & -4  \\  
\end{tabular}
\caption{List of candidate chiral ring generators for the $(A_1, D_4)$ theory (continued in Table \ref{tab:D4 ChiralOps2}). Here $R$, $r$, and $j$ are the IR $SU(2)_R$ Cartan, $U(1)_r$ charge, and left spin. $U(1)_B$ and $U(1)_m$ are $\CN=2$ flavor symmetries.}
    \label{tab:D4 ChiralOps1}
\end{table}
\begin{table}[]
    \centering
    \begin{tabular}{c c c c c c c c }
         Operators & $R$ & $r$ & $j$ & $U(1)_B$ & $U(1)_m$  \\
         $\phi q_1q_1\lambda_{\alpha}$ & 3/2 & 3/2 & 1/2 & 2 & 6 \\ 
         $\phi\tilde{q}_1q_1\lambda_{\alpha}$ & 3/2 & 3/2 & 1/2 & 0 & 0 \\ 
         $\phi\tilde{q}_1\tilde{q}_1\lambda_{\alpha}$ & 3/2 & 3/2 & 1/2 & -2 & -6 \\ 
         $\phi q_2q_2\lambda_{\alpha}$ & 3/2 & 1/2 & 1/2 & 2 & -2 \\ 
         $\phi\tilde{q}_2q_2\lambda_{\alpha}$ & 3/2 & 1/2 & 1/2 & 0 & 0 \\ 
         $\phi\tilde{q}_2\tilde{q}_2\lambda_{\alpha}$ & 3/2 & 1/2 & 1/2 & -2 & 2 \\ 
         $\phi\tilde{q}_2q_1\lambda_{\alpha}$ & 3/2 & 1 & 1/2 & 0 & 4 \\ 
         $\phi q_1q_2\lambda_{\alpha}$ & 3/2 & 1 & 1/2 & 2 & 2 \\ 
         $\phi\tilde{q}_1\tilde{q}_2\lambda_{\alpha}$ & 3/2 & 1 & 1/2 & -2 & -2 \\ 
         $\phi\tilde{q}_1q_2\lambda_{\alpha}$ & 3/2 & 1 & 1/2 & 0 & -4  \\ 
         $\phi\lambda_{\alpha}\lambda_{\beta}$ & 1 & 3/2 & 1 & 0 & 0 \\ 
         $q_1q_1\lambda_{\alpha}\lambda_{\beta}$ & 2 & 3/2 & 1 & 2 & 6 \\ 
         $\tilde{q}_1q_1\lambda_{\alpha}\lambda_{\beta}$ & 2 & 3/2 & 1 & 0 & 0 \\ 
         $\tilde{q}_1\tilde{q}_1\lambda_{\alpha}\lambda_{\beta}$ & 2 & 3/2 & 1 & -2 & -6 \\ 
         $q_2q_2\lambda_{\alpha}\lambda_{\beta}$ & 2 & 1/2 & 1 & 2 & -2 \\ 
         $\tilde{q}_2q_2\lambda_{\alpha}\lambda_{\beta}$ & 2 & 1/2 & 1 & 0 & 0 \\ 
         $\tilde{q}_2\tilde{q}_2\lambda_{\alpha}\lambda_{\beta}$ & 2 & 1/2 & 1 & -2 & 2 \\
         $\tilde{q}_2q_1\lambda_{\alpha}\lambda_{\beta}$ & 2 & 1 & 1 & 0 & 4 \\ 
         $q_1q_2\lambda_{\alpha}\lambda_{\beta}$ & 2 & 1 & 1 & 2 & 2 \\ 
         $\tilde{q}_1\tilde{q}_2\lambda_{\alpha}\lambda_{\beta}$ & 2 & 1 & 1 & -2 & -2\\ 
         $\tilde{q}_1q_2\lambda_{\alpha}\lambda_{\beta}$ & 2 & 1 & 1 & 0 & -4 \\ 
         $\lambda_{\alpha}\lambda_{\beta}\lambda_{\gamma}$ & 3/2 & 3/2 & 3/2 & 0 & 0 \\ 
     \end{tabular}
    \caption{Remaining candidate chiral ring generators for the $(A_1, D_4)$ theory (continued from Table \ref{tab:D4 ChiralOps1}). Here $R$, $r$, and $j$ are the IR $SU(2)_R$ Cartan, $U(1)_r$ charge, and left spin. $U(1)_B$ and $U(1)_m$ are $\CN=2$ flavor symmetries.}
    \label{tab:D4 ChiralOps2}
\end{table}

We can construct a list of naive multiplets using exactly the same set of procedures as in the previous subsection. Although the number of fields involved here is larger, we have done this in Tables \ref{tab:D4 ChiralOps1} and \ref{tab:D4 ChiralOps2}.

It will again prove useful to identify the $\bar\CE_{3/2}$ and $\hat\CB_1$ chiral operators. Unitarity implies that the primaries of the $\hat\CB_1$ multiplets and the primary and level-one descendant of the $\bar\CE_{3/2}$ multiplet cannot be composites built out of products of gauge-invariant operators (the level-two descendant of $\bar\CE_{3/2}$ can at most be built out of a product of two gauge invariant operators). Therefore, we can immediately identify the holomorphic moment maps
\begin{equation}\label{BhatlocA1D4}
\mu^a\in\left\{\beta_2~,\ \tilde q_2q_1~,\ q_1q_2~,\ \tilde q_1\tilde q_2~,\ \tilde q_1q_2~,\ \phi\tilde q_2\tilde q_2~,\ \phi\tilde q_2q_2~,\ \phi q_2q_2\right\}\ni \hat\CB_1^a~.
\end{equation}
For the $\bar\CE_{3/2}$ multiplet we have
\bea \label{ElocA1D4}
M_2\in (0,0)_{0,3/2} &\xrightarrow{Q^1_\alpha}& \phi\lambda_{\alpha}\in \left({1\over2},0\right)_{1/2,1}\xrightarrow{(Q^1)^2 }  \tilde q_1q_1\in(0,0)_{1,1/2}~.
\eea
We have mapped $\tilde q_1q_1$ to the level-two descendant of $\bar\CE_{3/2}$ using the fact that there is no other candidate built from a single generator in Table \ref{tab:A3 generators} or a product of two such generators that has the correct superconformal quantum numbers and is $SU(2)\times U(1)\subset SU(3)$ invariant (recall also that, by construction, $\phi^2$ decouples from the IR chiral ring).

In what follows, we will make use of the following superpotential constraints,
\begin{eqnarray}\label{A1D4Wconst}
\frac{\partial W}{\partial \beta_2}&=&\text{Tr}(\phi^2)=0~,\cr
\frac{\partial W}{\partial M_2}&=&\text{Tr}(\tilde{q}_2q_2)=0~,\cr
\frac{\partial W}{\partial\tilde{q}_1^a}&=&(\phi q_1)_a=0~,\cr 
\frac{\partial W}{\partial{q}_1^a}&=&_a(\tilde q_1\phi)=0~.
\end{eqnarray}

Let us now proceed to discuss potential $\bar\CB$ chiral generators of the theory
\begin{itemize}
\item Any operators with $r=j$ in Tables \ref{tab:D4 ChiralOps1} and \ref{tab:D4 ChiralOps2} cannot sit in $\bar\CB$ multiplets (see \eqref{Chiralloc2}). Since they are no $\bar\CD$ multiplets, these operators (modulo those in \eqref{BhatlocA1D4}) are trivial in the IR chiral ring.  This logic removes the following potential chiral ring generators
\begin{eqnarray}\label{rjgen}
&&\tilde q_2q_1\lambda_{\alpha}~,\ q_1q_2\lambda_{\alpha}~,\ \tilde q_1\tilde q_2\lambda_{\alpha}~,\ \tilde q_1q_2\lambda_{\alpha}~,\ \phi q_2q_2\lambda_{\alpha}~,\ \phi \tilde q_2q_2\lambda_{\alpha}~,\ \phi \tilde q_2\tilde q_2\lambda_{\alpha}~,\ \tilde q_2q_1\lambda_{\alpha}\lambda_{\beta}~,\cr&&q_1q_2\lambda_{\alpha}\lambda_{\beta}~,\ \tilde q_1\tilde q_2\lambda_{\alpha}\lambda_{\beta}~,\ \tilde q_1q_2\lambda_{\alpha}\lambda_{\beta}~,\ \lambda_{\alpha}\lambda_{\beta}\lambda_{\gamma}~.
\end{eqnarray}
\item Similar logic rules out operators with $r<j$. This reasoning removes the following potential chiral ring generators
\begin{equation}
\tilde q_2q_2~,\ q_2q_2\lambda_{\alpha}~,\ \tilde q_2q_2\lambda_{\alpha}~,\ \tilde q_2\tilde q_2\lambda_{\alpha}~,\ q_2q_2\lambda_{\alpha}\lambda_{\beta}~,\ q_2\tilde q_2\lambda_{\alpha}\lambda_{\beta}~,\ \tilde q_2\tilde q_2\lambda_{\alpha}\lambda_{\beta}~.
\end{equation}
\item Other operators are removed from the IR chiral ring via superpotential constraints. Indeed, constraints in \eqref{A1D4Wconst} remove
\begin{equation}
\phi q_1q_1~,\ \phi\tilde q_1q_1~,\ \phi\tilde q_1\tilde q_1~,\ \phi\tilde q_2q_1~,\ \phi q_1q_2~,\ \phi\tilde q_1\tilde q_2~,\ \phi\tilde q_1 q_2~.
\end{equation}
Note that there are no quantum corrections to these superpotenial removals since these are the unique operators with their superconformal and flavor quantum numbers.\footnote{This logic relies on the fact that $\phi^2$ decouples due to unitarity bound violations or equivalently, for this purpose, $\phi^2$ is set to zero in the IR chiral ring by a superpotential constraint.}
\item $\lambda^2$ has $(R,r(j,\bar{j}))=(1,1,(0,0))$. If it were a level-one descendant, the primary would have $(R,r(j,\bar{j}))=(1/2,3/2,(1/2,0))$. However, this would be a primary of a $\bar{\mathcal{D}}$ multiplet, which we know is absent.\footnote{Also, the primary cannot be constructed out of the list of chiral generators.} If it were a level-two descendant, the primary would have $(R,r(j,\bar{j}))=(0,2,(0,0))$, which would be an $\bar{\mathcal{E}}_2$ primary, which is again absent. Therefore, $\lambda^2$ must be trivial in the IR $\mathcal{N}=1$ chiral spectrum. 
\item $\phi q_1q_1\lambda_{\alpha}$, $\phi q_1\tilde q_1\lambda_{\alpha}$, and $\phi \tilde q_1\tilde q_1\lambda_{\alpha}$ have $(R,r(j,\bar{j}))=(3/2,3/2,(1/2,0))$. If they are level-one descendants, then the primary has $(R,r(j,\bar{j}))=(1,2,(1,0))$ (but then $r=1+j$, and we know there are no $\bar\CD$ multiplets) or $(R,r(j,\bar{j}))=(1,2,(0,0))$ (but such a primary cannot be constructed out of the list of $SU(2)$-neutral generators). If it is a level-two descendant, then the primary has $(R,r(j,\bar{j}))=(1/2,5/2,(1/2,0))$ ($M_2\phi\lambda_{\alpha}$ has these quantum numbers, but it is a descendant in an $\bar{\mathcal{E}}_{3}$ multiplet).
\item $\phi\lambda_{\alpha}\lambda_{\beta}$ has $(R,r(j,\bar{j}))=(1,3/2,(1,0))$. If it is a level-one descendant, then the primary has $(R,r(j,\bar{j}))=(1/2,2,(1/2,0))$ or  $(R,r(j,\bar{j}))=(1/2,2,(3/2,0))$ (but neither can be constructed from the list of generators). If it is a level-two descendant, then the primary has $(R,r(j,\bar{j}))=(0,5/2,(1,0))$ (which cannot be constructed from the list of generators since $\phi^2$ decouples).

\item $q_1q_1\lambda_{\alpha}\lambda_{\beta}$, $q_1\tilde q_1\lambda_{\alpha}\lambda_{\beta}$, and $\tilde q_1\tilde q_1\lambda_{\alpha}\lambda_{\beta}$ have $(R,r(j,\bar{j}))=(2,3/2,(1,0))$. If they are level-one descendants, then the primaries have $(R,r(j,\bar{j}))=(3/2,2,(1/2,0))$ (such operators cannot be constructed from the list of generators because of the vanishing of the operators \eqref{rjgen} in the IR chiral ring and the vanishing of $\phi^2$ in the IR theory) or $(R,r(j,\bar{j}))=(3/2,2,(3/2,0))$ (these operators also cannot be constructed from the list of generators). If it is a level-two descendant, then the primary has $(R,r(j,\bar{j})=(1,5/2,(1,0))$ (such an operator cannot be constructed from the list of generators since $\phi^2$ decouples).

\item $q_1q_1\lambda_{\alpha}$, $\tilde q_1q_1\lambda_{\alpha}$, and $\tilde q_1\tilde q_1\lambda_{\alpha}$ have $(R,r(j,\bar{j}))=(3/2,1,(1/2,0))$. These operators cannot be $\bar\CB$ superconformal primaries since $r<1+j$. If they are level-one descendants, then the primary has $(R,r(j,\bar{j}))=(1,3/2,(0,0))$ (such operators can potentially be constructed by taking a products of $M_2$ and one of the eight $\hat{\mathcal{B}}_1$ primaries; we will say more about this multiplet later) or $(R,r(j,\bar{j}))=(1,3/2,(1,0))$ (this is ruled out, as $r<1+j$, and because such operators cannot be constructed out of the list of generators). If it is a level two descendant, then the primary has $(R,r(j,\bar{j}))=(1/2,2,(1/2,0))$ (which cannot be constructed from list of generators).
\item $\phi \tilde q_2q_1\lambda_{\alpha}$, $\phi q_1q_2\lambda_{\alpha}$, $\phi \tilde q_1\tilde q_2\lambda_{\alpha}$, and $\phi \tilde q_1q_2\lambda_{\alpha}$ have $(R,r(j,\bar{j}))=(3/2,1,(1/2,0))$. These operators cannot be $\bar\CB$ superconformal primaries since $r<1+j$. If they are level-one descendants, then the primaries have $(R,r(j,\bar{j}))=(1,3/2,(1,0))$ (which fails due to $r<1+j$) or $(R,r(j,\bar{j}))=(1,3/2,(0,0))$ (such primaries can potentially be constructed by multiplying $M_2$ with a $\hat{\mathcal{B}}_1$ primary; we will say more about these operators later). If these operators are level-two descendants, then the primary has $(R,r(j,\bar{j}))=(1/2,2,(1/2,0))$ (but such primaries cannot be constructed from list of generators since $\phi^2$ decouples).
\end{itemize}
Therefore, in order to construct a $\bar\CB$ superconformal primary, we must build it out of products of chiral operators in the Coulomb branch multiplet $\bar{\CE}_{3/2}$ (i.e., $\{M_2,\phi\lambda_{\alpha},q_1\tilde{q}_1\}$) and the Higgs branch multiplets $\hat{\CB}_1$  (i.e., $\{q_1q_2,q_1\tilde{q}_2,\tilde{q}_1q_2,\tilde{q}_1\tilde{q}_2,\phi q_2q_2,\phi q_2\tilde{q}_2,\phi\tilde{q}_2\tilde{q}_2,\beta_2$\}). 

We can use the superpotential in \eqref{WA1D4} to constrain these products. To that end, we consider each quadratic product build out of the $\bar{\CE}_{3/2}$ and $\hat\CB_1$ chiral operators separately
\begin{itemize}
\item $M_2^2$ and $M_2\phi\lambda_{\alpha}$: These are the primary and the level-one descendant of the $\bar{\CE}_3$ multiplet.
\item $M_2q_1\tilde{q}_1$ and $(\phi\lambda_{\alpha})^2$: One linear combination is the level-two descendant of $\bar{\CE}_3$. The other independent linear combination, if nonzero, is a primary of a $\bar{\CB}$ multiplet. There is a unique candidate for the level-two descendant of this multiplet, $(q_1\tilde{q}_1)^2$. As we will show below, this operator vanishes in the IR chiral ring as a result of a superpotential constraint. Therefore, this $\bar{\CB}$ multiplet cannot exist in the $(A_1,D_4)$ theory.
\item $(q_1\tilde{q}_1)^2$ $(R,r(j,\bar j))=(2,1,(0,0))$: We have the following superpotential constraint
\begin{eqnarray}
\frac{\partial W}{\partial\phi^a}(q_1\tilde{q}_1)^a=0~,
\end{eqnarray}
which leads to,
\begin{equation}
2\beta_2\phi q_1\tilde{q}_1+\delta_{ab}(q_1\tilde{q}_1)^a(q_1\tilde{q}_1)^b=0~.
\end{equation}
We have already shown that $\phi q_1\tilde{q}_1$ is trivial in the IR $\CN=1$ chiral spectrum. Therefore, we are left with,
\begin{equation}
\delta_{ab}(q_1\tilde{q}_1)^a(q_1\tilde{q}_1)^b=0~.
\end{equation}
This constraint cannot receive quantum corrections since all other operators with the same superconformal quantum numbers have already been shown to vanish in the IR chiral ring.
\item $M_2 \hat\CB_1$ has $(R,r(j,\bar j))=(1,3/2(0,0))$: Here we consider the eight so-called \lq\lq mixed branch" primaries consisting of the product of the primaries in the Coulomb and Higgs branch chiral ring generators. The following chiral ring relation causes them to vanish:
\begin{equation}
\frac{\partial W}{\partial q_2}\phi q_2=M_2\phi q_2\tilde q_2=0~.
\end{equation}
By an $SU(3)$ flavor rotation, all other products of the form $M_2\hat\CB_1$ also vanish in the IR chiral ring. 
\item $\phi\lambda_{\alpha} \hat\CB_1$ has ($(R,r(j,\bar j))=(3/2,1(1/2,0))$): These eight operators cannot form a $\bar{\CB}$ primary since $r<1+j$.\footnote{In fact, we can show these operators are trivial in the IR chiral ring. Indeed, if they form a level-one descendant, then the primary has ($(R,r(j,\bar j))=(1,3/2(0,0))$, but this $\bar{\CB}$ multiplet has already been removed by the superpotential constraint above. If they form a level-two descendant, then the primary has $(R,r(j,\bar j))=(1/2,2(1/2,0))$, and the only candidate operator is $M_2\phi\lambda_{\alpha}$. However, this is itself a level-one descendant.\label{full1}}
\item $q_1\tilde q_1\hat\CB_1$ has ($(R,r(j,\bar j))=(2,1/2(0,0))$: These operators cannot be $\bar{\CB}$ primaries since $r<1$.\footnote{In fact, these operators are trivial in the IR chiral ring. Indeed, if they form a level-one descendant, then the primary has $(R,r(j,\bar j))=(3/2,1(1/2,0))$, but this also has $r<1+j$. If it is a level-two descendant, then the primary has ($(R,r(j,\bar j))=(1,3/2(0,0))$, but we have already eliminated this $\bar{\CB}$ multiplet by the superpotential constraint.\label{full2}}
\end{itemize}
Therefore, we see that we cannot construct a $\bar\CB$ primary, and so there are no $\bar\CB$ multiplets in the $(A_1, D_4)$ SCFT.\footnote{In fact, using the argument in footnotes \ref{full1} and \ref{full2}, we directly see that all operators except those in the Coulomb branch and Higgs branch chiral rings vanish in the IR FCS.}
In particular, we see that
\begin{equation}
{\rm FCS}_{(A_1, D_4)}=\langle \bar\CE_{3/2}, \hat\CB_1\rangle/I~, \ \ \ \bar\CB\not\in {\rm FCS}_{(A_1, D_4)}~,
\end{equation}
where $I$ is the Joseph ideal constraint.

\subsec{$SU(2)$ SQCD with $N_f=4$}
Finally, we discuss the $\CN=2$ Lagrangian theory of $SU(2)$ SQCD with four flavors. Unlike the previous cases, this theory is not isolated (it has an exactly marginal coupling).

In the interacting theory, we can again show that all chiral ring generators are in the Coulomb branch and Higgs branch chiral subrings.\footnote{By the interacting theory, we mean the interacting theory at generic points on the conformal manifold.} In this case, this means the chiral ring generators live in $\bar{\CE}_2$ or $\hat{\CB}_1^M$ multiplets (with $M$ an $SO(8)$ adjoint index). Since the gauge group is $SU(2)$, the naive list of chiral generators is very similar to the lists in the Argyres-Douglas examples we treated before (which were based on $SU(2)$ gauge theory Lagrangians), so we will keep the discussion brief. 

\begin{table}[]
    \centering
    \begin{tabular}{c c c c c c}
         Field & R & r & j & $SU(2)_{\text{gauge}}$ & $SO(8)$ \\
         $\phi$ & 0 & 1 & 0 & {\rm adj.} & 1 \\ 
         $\lambda_{\alpha}$ & 1/2 & 1/2 & 1/2 & {\rm adj.} & 1 \\
         $Q^a$ & 1/2 & 0 & 0 & 2 & 8 \\          
    \end{tabular}
    \caption{List of chiral fields appearing in the Lagrangian of the $SU(2)$ theory with four fundamental flavors. Here $r$, $R$, and $j$ are the $U(1)_r$, $SU(2)_R$ Cartan, and left spin respectively. In the rightmost column we record the representation under the $SO(8)$ flavor symmetry (the matter fields transform in the vector representation). If we want to write our matter field charges in terms of $SU(4)\subset SO(8)$, we can define $q^i:=Q^i$ and $\tilde q^i:=Q^{i+4}$ where $i=1,\cdots,4$.}
    \label{tab: D4-Fields}
\end{table}

We use Table \ref{tab: D4-Fields} to write down the list of UV chiral fields. The Lagrangian in this case is
\begin{equation}\label{Nf4W}
W=\tau \phi(Q^aQ_a)~.
\end{equation}
Note that in terms of $SU(4)\subset SO(8)$, we can define $q^i:=Q^i$ and $\tilde q^i:=Q^{i+4}$ where $i=1,\cdots,4$.

The naive chiral ring generators are the following:
\begin{itemize}
\item $\phi^2$, $\phi\lambda_{\alpha}$, $\lambda^2$: These are, respectively, the primary, the level one, and the level-two descendants of the $\bar{\CE}_2$ multiplet. In the interacting theory, $\lambda^2$ mixes with $\phi(Q^aQ_a)$ via \eqref{Nf4W}.
\item $M^{ab}=Q^{[a}Q^{b]}$ (there are 28 of these operators transforming in the adjoint of $SO(8)$). These are the $\hat\CB_1^A$ Higgs branch generators housing the Noether currents of the flavor symmetry.
\item The other possible operators with $j=0$ have the following form: $\phi QQ$ ($R=1$, $r=1$). If they are non-trivial in the chiral ring, these operators can only sit in $\bar\CD$ multiplets. In the interacting theory, one linear combination with $\lambda^2$ becomes the level-two descendant of $\bar\CE_2$.
\item At $j=1/2$, the possible operators have either of two forms, (1) $QQ\lambda$ ($R=3/2$, $r=1/2$). If they are non-trivial in the chiral ring, these operators can only be descendants of $\bar{\CD}_{1(0,0)}$ multiplets, or (2) $\phi QQ\lambda$ ($R=3/2$, $r=3/2$). These operators can be primaries of $\bar{\CD}_{\frac{3}{2},(\frac{1}{2},0)}$ multiplets or descendants of $\bar{\CB}_{1,2(0,0)}$, $\bar\CD_{1,(1,0)}$, or $\bar{\CB}_{\frac{1}{2},\frac{5}{2}(\frac{1}{2},0)}$ multiplets.
\item At $j=1$, the possible generators are of the following forms, (1) $\phi\lambda\lambda$ ($R=1$, $r=2$). These can be a primary of a $\bar{\CD}_{1(1,0)}$ multiplet or descendants of $\bar{\CB}_{\frac{1}{2},\frac{5}{2}(\frac{1}{2},0)}$ or $\bar\CD_{{1\over2}({3\over2},0)}$, or (2) $QQ\lambda\lambda$ ($R=2$, $r=1$). These can only be descendants of $\bar{\CD}_{\frac{3}{2},(\frac{1}{2},0)}$ or $\bar{\CB}_{1,2(0,0)}$ multiplets.
\item At $j=3/2$, the only possible generator is of the form $\lambda\lambda\lambda$ ($R=r=3/2$), which can only be the descendant of a $\bar{\CD}_{1(1,0)}$ multiplet.
\end{itemize}

To summarize, we see that these operators may lie in the following multiplets: {\bf(1)} $\bar{\CD}$ multiplets. However, in the interacting theory, these multiplets are not present due to the same logic as in the $(A_1, A_3)$ and $(A_1, D_4)$ cases: the associated chiral algebra is of AKM type ($\widehat{so(8)}_{-2}$ in this case \cite{Beem:2013sza}). To see this statement more concretely, note that, when we turn on interactions, thirty six linear combinations of the $\phi QQ$ and $\lambda^2$ operators discussed above pair up with thirty-six of the thirty-seven stress tensor multiplets in the free theory to become long multiplets (the remaining stress tensor multiplet is protected along the full conformal manifold; see \cite{Buican:2016arp} for further details of this argument). {\bf(2)} As descendants in a $\bar{\CB}_{\frac{1}{2},\frac{5}{2}(\frac{1}{2},0)}$ multiplet. However, we cannot construct a candidate for the superconformal primary of this multiplet in a rank-one theory. {\bf(3)} As a descendant in a $\bar{\CB}_{1,2(0,0)}$ multiplet (we can construct a candidate primary for this multiplet, but the multiplet itself is known to be absent in this theory since it does not have a mixed branch; more on this later). Therefore, we conclude that, in the interacting theory, the only non-trivial chiral generators lie in $\bar{\CE}_2$ or $\hat{\CB}_1$ multiplets.

Once again, we proceed to take normal-ordered quadratic products of the above Coulomb and Higgs branch generators. There are two possible sources of $\bar{\CB}$ multiplets which we can form by taking products of the above generators. We now study each case individually.

The Coulomb branch multiplet, $\bar{\CE}_2$, has the following chiral operators (for simplicity, we define $M_2$ at $\tau=0$),
\begin{equation}
M_0:=\phi^2~,\ M_{1\alpha}:=\phi\lambda_{\alpha}~,\ M_2=\lambda^2~.
\end{equation}
Since the theory is rank one, there is no candidate for a superconformal primary of a $\bar{\CB}_{1/2,3/2,(1/2,0)}$ multiplet. 

We can take a linear combination of these operators
\begin{equation}\label{B1neutral}
\CO:=    M_0M_2+\kappa M_1^2~,
\end{equation}
where $\kappa\in\mathbb{C}$, and $\CO$ is a superconformal primary of a $\bar{\CB}_{1,3(0,0)}$ multiplet. It is trivial to check that this operator is present in the free theory. In the index, it is also easy to check that there is no $\bar{\CB}_{1,3(0,0)}$ contribution at the leading order it can appear. The reason is that there are thirty-eight $\bar\CC_{0,2(0,0)}$ multiplets in the free theory: thirty-seven from $\bar\CE_2\times\hat\CC_{0(0,0)}$ and one of the form $f_{ABC}\epsilon^{\alpha\beta}\epsilon_{IJ}\lambda_{\alpha}^{IA}\lambda_{\beta}^{JB}\phi^C$, where we have contracted $SU(2)_R$ and Lorentz indices of the gauginos (note that the anti-symmetrization of gauge indices makes this latter operator a superconformal primary). On the other hand, there are thirty-seven $\bar{\CB}_{1,3(0,0)}$ multiplets: thirty-six arising from $\bar\CE_2\times\bar\CD_{1(0,0)}$ and one linear combination as in \eqref{B1neutral}. The corresponding index contributions cancel up to a net $\bar\CC_{0,2(0,0)}$ contribution (note that three of the $\bar\CC_{0,2(0,0)}$ multiplets are flavor singlets and so are two of the $\bar{\CB}_{1,3(0,0)}$ contributions).

At leading order in the gauge coupling (i.e., at leading order in $\tau$), it is easy to see that two flavor-neutral $\bar{\mathcal{C}}_{0,2(0,0)}$ multiplets and the $\bar{\CB}_{1,3(0,0)}$ multiplet in \eqref{B1neutral} remain as short multiplets (unlike other $\bar{\mathcal{C}}_{0,2(0,0)}$ and $\bar{\CB}_{1,3(0,0)}$ pairs in the theory that become long multiplets at leading order when we turn on the superpotential in \eqref{Nf4W}). The $\bar{\CB}_{1,3(0,0)}$ multiplet is not protected from recombination at higher orders in the coupling (but there is always a protected $\bar{\mathcal{C}}_{0,2(0,0)}$ multiplet).

Finally, we can form a potential primary of a $\bar{\CB}_{1,2(0,0)}$ multiplet from the operator,
\begin{equation}
M_0\text{Tr}(QQ)~,
\end{equation}
which would correspond to a mixed branch. These operators vanish in the chiral ring at leading order in the coupling (they pair up with an adjoint-valued $\bar\CC_{0,1(0,0)}$ multiplet to become long multiplet descendants). This statement is consistent with the fact that the theory does not have a mixed branch. In fact, any such candidate obtained from the $\bar{\CE}_{2m}\times\hat{\CB}_n$ OPE for $m,n\in\mathbb{N}$ will not correspond to a $\bar{\CB}_{n,2m(0,0)}$ multiplet for the same reason.

\section{Conclusion}
In this paper, we have explored various new roles that $\bar\CB$ multiplets play in 4d $\CN=2$ SCFTs. We have shown that these operators are ubiquitous and are connected to many interesting phenomena. Our work also raises several questions:
\begin{itemize}
\item Can our algebraic proof that the product of primaries in $\bar\CD_{1/2(0,0)}\times\bar\CD_{1/2(0,0)}$ is non-vanishing and produces a $\bar\CB$ multiplet be generalized to show that any $\CN>2$ theory has an infinite chiral ring generated by the primary of $\bar\CD_{1/2(0,0)}$?\footnote{Such a ring would contain an infinite number of $\bar\CB$ multiplets.} Since this multiplet houses the extra supercurrents in its higher components, it would be interesting to understand if such an infinite chiral ring exists as a consequence of physics related to Ward identities for extended SUSY. If so, this may be a step in an abstract CFT-based proof that all $\CN>2$ theories have a mixed branch of moduli space.
\item We saw that the existence of (adjoint valued) $\bar\CB$ multiplets can be a diagnostic of the $\mathbb{Z}_2$-valued $Sp(N)$ 't Hooft anomaly in \cite{Witten:1982fp}. What about more general 't Hooft anomalies? Can these operators help diagnose the existence of 2-groups and other more elaborate structures?
\item We saw that $\bar\CB$ multiplets form a natural ideal in the FCS ring. Can we use this fact to prove new results about chiral rings in 4d $\CN=2$ SCFTs?
\item The $\bar\CB$ production channels we discussed do not lead to $\bar\CB$ chiral ring generators. It would be interesting to understand the most general conditions under which $\bar\CB$ multiplets can and cannot house chiral ring generators (see footnote \ref{gauging}). Can gauging discrete symmetries help?
\item We conjectured a bound on the quantum numbers of $\bar\CB$ multiplets. It would be interesting to prove (or disprove) this bound.
\item We have seen that the particular isolated rank-one theories we studied do not have $\bar\CB$ multiplets (of course, rank one theories that violate some of the criteria we imposed do have $\bar\CB$ mulitplets; examples include $SU(2)$ $\CN=4$ SYM among others). It would be interesting to better understand why this is the case (and to extend our analysis to other cases with $\CN=1$ Lagrangians). Our analysis relied on the existence of certain $\CN=1$ Lagrangians that flow to $\CN=2$ in the IR. Can we give an abstract argument, perhaps using properties of 't Hooft anomalies and $a$-maximization, that any rank-one IR SCFT of the general type we studied arising from an $\CN=1$ Lagrangian has no $\bar\CB$ multiplets? Can we extend our analysis to other rank-one theories even when there is no known Lagrangian? Can we give a more manifest and dynamical proof that the rank-one theories we studied do not have $\bar\CB$ multiplets?
\end{itemize}
We hope to return to some of these questions soon.

\ack{We are grateful to C.~Bhargava, H.~Jiang, A.~Manenti, and T.~Nishinaka for comments, discussions, and collaboration on related work. A.~B. and M.~B. were partially supported by the grant “Relations, Transformations, and Emergence in Quantum Field Theory” from the Royal Society. M.~B. was also supported by the grant “Amplitudes, Strings and Duality” from STFC. No new data were generated or analysed during this study.}

\newpage
\bibliography{chetdocbib}

\end{document}